\documentclass[preprint,3p,12pt]{elsarticle}
\usepackage{mathrsfs}
\usepackage{amsmath}
\usepackage{stmaryrd}
\usepackage{bbding}
\usepackage{dcolumn}
\usepackage{graphicx}
\usepackage{amsfonts}
\usepackage{amssymb}
\usepackage{psfrag}
\usepackage{wrapfig}
\usepackage{subfigure}
\usepackage{makeidx}
\usepackage{bm}
\usepackage{epsf}
\usepackage{epsfig}
\usepackage{setspace}
\usepackage{graphicx}
\usepackage{epstopdf}
\usepackage{psfrag}
\usepackage{subfigure}
\usepackage{color}
\usepackage{comment}
\usepackage{algorithm}
\usepackage{algorithmic}

\newcommand \td {\mathrm{~d}}
\begin{document}

\title{A high-order compact gas-kinetic scheme in a rotating coordinate frame and on sliding mesh}

\author[HKUST1]{Yue Zhang}
\ead{yzhangnl@connect.ust.hk}	
\author[HKUST1]{Xing Ji}
\ead{xjiad@connect.ust.hk}
\author[HKUST1,HKUST2,HKUST3]{Kun Xu\corref{cor1}}
\ead{makxu@ust.hk}

\address[HKUST1]{Department of Mathematics, Hong Kong University of Science and Technology, Clear Water Bay, Kowloon, Hong Kong}
\address[HKUST2]{Department of Mechanical and Aerospace Engineering, Hong Kong University of Science and Technology, Clear Water Bay, Kowloon, Hong Kong}
\address[HKUST3]{Shenzhen Research Institute, Hong Kong University of Science and Technology, Shenzhen, China}
\cortext[cor1]{Corresponding author}

\begin{abstract}
This paper extends the high-order compact gas-kinetic scheme (CGKS) to compressible flow simulations on a rotating coordinate frame. 
The kinetic equation with the inclusion of centrifugal and Coriolis acceleration is used in the construction of the scheme. 
With the updates of both cell averaged conservative variables and their gradients in the rotating and stationary domains, a third-order compact reconstruction is developed with sliding interface between them. 
To properly capture shock wave and complicated wave interaction, the HWENO-type non-linear reconstruction and gradient compression factors are incorporated in the scheme. For achieving high-order time accuracy, based on the flux function and its time derivative the multi-stage multi-derivative (MSMD) time stepping method is implemented in the scheme for the fourth-order accuracy with two stages.
The CGKS is validated by many test cases from subsonic acoustic wave propagation to the high Mach number shock interaction in a rotating frame. The compact scheme achieves high-order accuracy and remarkable robustness.
\end{abstract}


\maketitle

\section{Introduction}

Flow simulations with rotating parts, including turbo-machinery, helicopters, tilt-rotors, and ship propellers, have significant industrial applications. The computational domain is usually divided into moving and stationary parts with a sliding interface between them.
This paper is about the development of a high-order compact gas-kinetic scheme on a rotating coordinate frame and connect its solution with
the stationary domain through a sliding interface.

The gas-kinetic scheme (GKS) is a kinetic theory-based numerical method to solve the Euler and Navier-Stokes equations \cite{xuGKS2001}.
Under the initial condition of a generalized Riemann problem, a time accurate gas distribution function is constructed in GKS
 to calculate the numerical fluxes and evaluate the time-dependent flow variables at a cell interface.
As a result, both cell averaged flow variables and their gradients can be updated.
Therefore, the HWENO-type method and the
 two-step multi-resolution WENO reconstruction
can be developed in the scheme for the high-order spatial data reconstruction \cite{zhuNewFiniteVolume2018,jiMultiresolution2021}.
At the same time, due to the time accurate flux function, the multi-stage multi-derivative (MSMD) method can be used to update the solution with  high-order temporary accuracy. Specifically, the two-stage fourth-order (S2O4) time stepping method is used in the compact GKS (CGKS) \cite{liTwoStageFourthOrder2016,li2019two}.
The CGKS has been constructed on both structured and unstructured meshes in 2D and 3D cases  \cite{ji4order2018,jiThreedimensionalCompactHighorder2020,
zhaoCompactHigherorderGaskinetic2019,zhaoAcousticShockWave2020,zhaoCompactHighorderGaskinetic2022a}.
In order to further improve the robustness of the scheme in high-speed flow simulations,
the following modeling has been further incorporated in the scheme.
First, the evolution of possible discontinuous flow variables at different sides of a cell interface are constructed
for updating reliable cell averaged gradient of flow variables \cite{zhaoDirectModelingComputational2021}.
Second, the nonlinear limiting process is implemented on the high-order time derivative of the flux function under MSMD framework.
Equipped with the above remedies, the CGKS on 3D unstructured mesh is extremely robust in hypersonic flow computation and
a large time step, such as CFL number $\approx 0.8$, can be used in the fourth-order compact scheme.
Alternatively, a gradient compression factor is designed to improve the robustness and efficiency of CGKS in case of low quality mesh \cite{jiGC2021}.

There are two ways to solve the flow problems with a rotating mesh movement.
The first one is the arbitrary Lagrangian-Eulerian (ALE)-based moving mesh method \cite{hirt,zhang2015,duan2020}.
These methods are similar to the methods under the unified coordinates \cite{hui,jin}, where flow variables and geometric conservation laws have
to be solved simultaneously.
The another approach fixes the coordinate on a rotating frame.
In such a non-inertia reference of frame, the centrifugal and Coriolis forces will appear in the kinetic governing equation \cite{zhouThreeDimensionalGasKineticBGK2019}.
In this paper, we are going to develop the GKS by following the second approach.
With the inclusion of external forces in the kinetic equation, the CGKS can be constructed with the inclusion of forcing effect on the particle trajectory \cite{xushallow_water2002b, luo2011}.
The corresponding macroscopic governing equations solved by the CGKS will be derived using the Chapman-Enskog expansion.

The sliding-mesh method has been developed for many years.
Johnstone et al. \cite{johnstone2015} proposed a novel sliding-mesh method based on a characteristic interface condition.
Their new sliding gird technique requires only a single layer of halo nodes in the communication process.
Luis Ramírez et al. \cite{ramirez2015} developed a high-order sliding mesh interface to simulate unsteady viscous flow.
Both compressible inviscid flow and incompressible viscous flow were simulated with the
moving least squares (MLS) \cite{MLS} reconstruction at the sliding interface.
All above sliding mesh methods are for the non-compact schemes.
The high-order compact schemes have much advantages in comparison with non-compact ones due to the
compact stencils around the sliding interface.
Many high-order compact schemes have been developed based on evolution of cell's inner degrees of freedom, such as discontinuous Galerkin (DG) \cite{DG}, spectral difference (SD) \cite{SD}, flux reconstruction (FR) \cite{huynh2007}, and correction procedure via reconstruction (CPR)\cite{CPR2011}. Based on FR/CPR scheme, Duan et al.\cite{duan2020} developed a sliding mesh method by using an auxiliary Cartesian grid to exchange information between the sliding interface. Based on the SD method, Zhang and Liang \cite{zhang2015} used mortar elements to project flow variables and fluxes back and forth. To improve the adaptability of geometry, Zhang et al. \cite{zhang2018flux} extended their work to deal with arbitrarily non-uniform mesh for the FR method. Recently, Gao \cite{gao2022} developed a three-dimension sliding mesh method based on the mortar approach and  applied it to the turbine and noise problem. By adopting Sutherland-Hodgman algorithm \cite{Sutherland197432}, the polygon clipping method can be used to deal with complicated geometry.
In the current study, a third-order finite volume CGKS only requires Neumann neighboring cells in the reconstruction.
Therefore, it becomes straightforward to construct the corresponding CGKS with sliding mesh.
Here, a ghost cell will be created by merging several cells in the reconstruction around the sliding interface.
To ensure the conservative property, the mortar interface is  generated for the  calculation of fluxes.

This paper is organized as follows. The kinetic BGK equation and GKS in a rotating coordinate frame will be introduced in section 2.
Section 3 is about the two-stage four-order time integrating method for the solution updates with source terms.
Section 4 concentrates on the initial reconstruction.
The treatment of the sliding interface is presented in section 5. Many test cases will be used  to validate the current method in section 6.
The last section is the conclusion.

\section{Gas-Kinetic Scheme}
\subsection{BGK equation in rotating framework}
The gas-kinetic BGK equation in a rotating frame is
\begin{equation*}
  \frac{\partial f}{ \partial t}+ \boldsymbol{w}\cdot \nabla_x f + \boldsymbol{a}_w\cdot \nabla_w f = \frac{g-f}{\tau},
\end{equation*}
where $f=f(\boldsymbol{x},t,\boldsymbol{w},\xi)$ is the gas distribution function, $g$ is the corresponding equilibrium state, and $\tau$ is
the collision time. $\boldsymbol{w}=(w_1,w_2,w_3)$ is the particle velocity in the rotating frame. And the acceleration in the rotating frame is
\begin{equation*}
  \boldsymbol{a}_w =\frac{\td\boldsymbol{w}}{\td t} = -\boldsymbol{\Omega}\times(\boldsymbol{\Omega}\times \boldsymbol{r})-2(\boldsymbol{\Omega}\times \boldsymbol{w}),
\end{equation*}
where $\boldsymbol{\Omega}$ is the angular velocity of the rotating frame, and $\boldsymbol{r}$ is a position vector from the origin of rotation to the position of the particle. $-\boldsymbol{\Omega}\times(\boldsymbol{\Omega}\times \boldsymbol{r})$ is centrifugal force and $-2(\boldsymbol{\Omega}\times \boldsymbol{w})$ is Coriolis force.    Denote $\boldsymbol{v}=(v_1,v_2,v_3)$ as the particle velocity
in absolute inertia reference of frame, the relationship among velocities is
\begin{equation*}
  \boldsymbol{v} =\boldsymbol{U}+\boldsymbol{w},
\end{equation*}
where $\boldsymbol{U}=\boldsymbol{\Omega}\times\boldsymbol{r}$  is convection velocity due to the frame rotation.
According $\boldsymbol{a}_v= \frac{\td\boldsymbol{v}}{\td t} = \frac{\td\boldsymbol{w}}{\td t}+ (\boldsymbol{\Omega}\times \boldsymbol{w}) $, the acceleration term in the rotating frame can be expressed as $-(\boldsymbol{\Omega}\times \boldsymbol{v}) \cdot\nabla_v f $. Then the BGK equation becomes
\begin{equation}\label{BGK}
  \frac{\partial f}{ \partial t}+ \boldsymbol{w}\cdot \nabla_x f - (\boldsymbol{\Omega}\times \boldsymbol{v})\cdot\nabla_v f = \frac{g-f}{\tau},
\end{equation}
where $f$ can be defined by absolute velocity $\boldsymbol{v}$, such as $f=f(\boldsymbol{x},t,\boldsymbol{v},\xi)$.
The collision term in the above equation describes the evolution process from a non-equilibrium state to an equilibrium one with
the satisfaction of compatibility condition
\begin{equation*}
  \int \frac{g-f}{\tau} \Psi \td\boldsymbol{v}\td\Xi = 0,
\end{equation*}
where $\Psi=(1,v_1,v_2,v_3,\frac{1}{2}(v_1^2+v_2^2+v_3^2+\xi^2))^T$ and $\td\Xi=\td\xi_1\cdots\td\xi_K$ ($K$ is the number of internal degree of freedom, i.e. $K=2$ for three-dimensional diatomic gas). Based on the Chapman-Enskog Expansion (see \ref{C-E}),
the Euler and N-S equations in the rotating frame can be obtained. The N-S equations in a rotating frame are
\begin{equation*}
  \begin{aligned}
    \frac{\partial \rho }{\partial t} &+\nabla\cdot\rho(\boldsymbol{V}-\boldsymbol{U})=0,\\
    \frac{\partial \rho \boldsymbol{V}}{\partial t} &+\nabla\cdot[\rho (\boldsymbol{V}-\boldsymbol{U}) \boldsymbol{V}+pI -\overline{\overline{\sigma}}]=-\boldsymbol{\Omega}\times \rho\boldsymbol{V},\\
    \frac{\partial \rho E }{\partial t} &+\nabla\cdot[\rho H(\boldsymbol{V}-\boldsymbol{U})+p\boldsymbol{V} -\kappa \nabla T -\overline{\overline{\sigma}}\cdot \boldsymbol{V}]=0,\\
  \end{aligned}
\end{equation*}
where $\rho,\boldsymbol{V},p,T,E,H$ and $\overline{\overline{\sigma}}$ are the density, absolute velocity, pressure, temperature, energy, enthalpy and viscosity stress of fluid. With the gradient of temperature $\nabla T$ and viscosity stress $\overline{\overline{\sigma}}$ equal to zero, the N-S equations become Euler Equations.

  \subsection{Finite volume method}
  The whole domain $\Omega$ is discretized into small cells $\Omega_i$
  \begin{equation*}
    \Omega =\bigcup \Omega_i,\ \Omega_i \bigcap \Omega_j=\phi (i \neq j).
  \end{equation*}
  The boundary can be expressed as
  \begin{equation*}
    \partial \Omega_i=\bigcup_{p=1}^{N_f}\Gamma_{ip}.
  \end{equation*}
  Taking moments of the BGK equation (\ref{BGK}) and integrating over the cell $\Omega_i$, the semi-discretized form of the finite volume scheme can be written as
\begin{equation}\label{fvm}
  \frac{\td \boldsymbol{W_i}}{\td t} = -\frac{1}{|\Omega_i|}\sum_{p=1}^{N_f}\int_{\Gamma_{ip}} \boldsymbol{F}(\boldsymbol{W}) \cdot \boldsymbol{n}_p dS + \boldsymbol{S}(\boldsymbol{W})
  := \mathcal{L}_F\left(\boldsymbol{W}\right) + \boldsymbol{S}(\boldsymbol{W}),
\end{equation}
where $\boldsymbol{W}_i$ is the cell average conservative value, $|\Omega_i|$ is the volume of cell $\Omega_i$, $\boldsymbol{F}$ is the flux via cell surface, $\boldsymbol{n}_p=(n_1,n_2,n_3)^T$ is the normal direction of cell surface and $\boldsymbol{S}$ is the source term due to rotation. The integration of flux can be approximated by Gaussian integrating (the index $i$ is omitted )
\begin{equation*}
  \int_{\Gamma_{p}} \boldsymbol{F}(\boldsymbol{W}) \cdot \boldsymbol{n}_p dS  \approx |S_p|\sum_{k=1}^{M_p} \omega_k \boldsymbol{F}(\boldsymbol{x}_{p,k},t)\cdot \boldsymbol{n}_{p,k},
\end{equation*}
where $|S_p|$ is the cell surface area, $\omega_k$ is the weight of Gaussian integrating, and $\boldsymbol{x}_{p,k}$ is the position of Gaussian points on the cell surface. To calculate the flux through the surface, we can use coordinate transform
\begin{equation*}
  \boldsymbol{F}(\boldsymbol{x}_{p,k},t)\cdot \boldsymbol{n}_p = \mathbf{T}^{-1}\widetilde{\boldsymbol{F}}(\mathbf{T}\boldsymbol{W}) = \mathbf{T}^{-1}\widetilde{\boldsymbol{F}}(\widetilde{\boldsymbol{W}}),
\end{equation*}
where $\mathbf{T} = diag(1,\mathbf{T}^{\prime},1)$ is rotating matrix, and
\begin{equation*}
  \mathbf{T}^{\prime}=\left(\begin{array}{ccccc}
   n_{1} & n_{2} & n_{3}\\
   -n_{2} & n_{1}+\frac{n_{3}^{2}}{1+n_{1}} & -\frac{n_{2} n_{3}}{1+n_{1}} \\
 -n_{3} & -\frac{n_{2} n_{3}}{1+n_{1}} & 1-\frac{n_{3}^{2}}{1+n_{1}}
  \end{array}\right), \quad n_{1} \neq-1,
  \end{equation*}
and when $n_1=-1$, $\mathbf{T}^\prime$ becomes $diag(-1,-1,1)$. And the flux can be evaluated by
  \begin{equation}\label{flux}
    \widetilde{\boldsymbol{F}}=\int f(\tilde{\boldsymbol{x}}_{p,k},t,\tilde{\boldsymbol{v}},\xi)\tilde{w}_1 \widetilde{\boldsymbol{ \Psi}} \td\boldsymbol{v}\td\Xi,
  \end{equation}
where the origin point of the local coordinate is $\tilde{\boldsymbol{x}}_{p,k}=(0,0,0)$ with x-direction in $\boldsymbol{n}_p$, and $\widetilde{\boldsymbol{ \Psi}} =(1,\tilde{v}_1,\tilde{v}_2,\tilde{v}_3,\frac{1}{2}(\tilde{v}_1^2+\tilde{v}_2^2+\tilde{v}_3^2+\xi^2))^T$. The microscopic velocities in local coordinate are given by $\tilde{\boldsymbol{v}} = \mathbf{T}^\prime \boldsymbol{v}$  and $\tilde{w}_1= n_1w_1 +n_2w_2+n_3w_3$.

  \subsection{Gas evolution model}
  In order to construct the numerical fluxes at $\boldsymbol{x} = (0,0,0)^T$, the integral solution of the BGK equation Eq.(\ref{BGK}) is
  \begin{equation}\label{solbgk}
    f(\boldsymbol{x},t, \boldsymbol{v},\xi)=\frac{1}{\tau_n}\int_0^t g(\boldsymbol{x}^{\prime},t^{\prime}\boldsymbol{v}^{\prime},\xi)e^{-(t-t^{\prime})/\tau_n}\td t^{\prime}+ e^{t/\tau_n}f_0(\boldsymbol{x}_{0},\boldsymbol{v}_0),
  \end{equation}
  where the partial absolute velocity is
  \begin{equation*}
  \begin{aligned}
  \boldsymbol{v}&=\boldsymbol{v}^{\prime}+\int_{t^{\prime}}^t\boldsymbol{a}_v \td \tilde{t} \\
 & = \boldsymbol{v}^{\prime}+\left[\boldsymbol{v}^{\prime} -\left(\boldsymbol{v}^{\prime}  \cdot \frac{\boldsymbol{\Omega}}{\Omega}\right) \frac{\boldsymbol{\Omega}}{\Omega}\right](1-\cos{\Omega(t-t^\prime)}) - \left(  \frac{\boldsymbol{\Omega}}{\Omega}\times\boldsymbol{v}^{\prime}  \right)\sin{\Omega(t-t^\prime) } \\
 & \approx \boldsymbol{v}^{\prime}-(\boldsymbol{\Omega}\times\boldsymbol{v}^{\prime}) (t-t^\prime).
  \end{aligned}
  \end{equation*}
 In the rotating frame, the particle velocity and trajectory become
  $\boldsymbol{w}^{\prime}-(\boldsymbol{\Omega}\times\boldsymbol{v}^{\prime}) (t-t^\prime)$ and
  \begin{equation*}
    \boldsymbol{x}\approx\boldsymbol{x}^{\prime}+\boldsymbol{w}^\prime(t-t^{\prime})+ \frac{1}{2} (\boldsymbol{\Omega}\times\boldsymbol{v}^{\prime}) (t-t^\prime)^2 .
  \end{equation*}
In Eq.(\ref{solbgk}), $f_0$ is the initial gas distribution function, and $g$ is the corresponding equilibrium state. $(\boldsymbol{x}_0,\boldsymbol{v}_0)$ are the initial position and velocity by tracing back particles $(\boldsymbol{x},\boldsymbol{v})$
at time $t$ back to $t=0$.
$\tau_n$ is the numerical collision time \cite{luo2013}. For inviscid flow, it set as 
$$\tau_n=C_1\Delta t + C_2\frac{|p_l-p_r|}{p_l+p_r}\Delta t ,$$ 
and for viscous flow, it is
$$\tau_n=\tau + C_2\frac{|p_l-p_r|}{p_l+p_r}\Delta t .$$
In this paper, we have $C_1=0.01,C_2=5.0$.

Before the construction of the initial distribution function $f_0$ and equilibrium state $g$, we first denote
  \begin{equation*}
    \boldsymbol{a}=(a_1,a_2,a_3) = \nabla_x g/g, A = g_t/g.
  \end{equation*}
In the following derivation, quadratic terms of time will be ignored directly.
With the Consideration of possible discontinuity at an interface, the initial distribution is constructed as
  \begin{equation}\label{ini_gas}
    f_0(\boldsymbol{x}_{0},\boldsymbol{v}_0)=f_0^l(\boldsymbol{x}_{0},\boldsymbol{v}_0) (1 -\mathbb{H}(x_1)) +f_0^r(\boldsymbol{x}_{0},\boldsymbol{v}_0)\mathbb{H}(x_1),
  \end{equation}
  where $\mathbb{H}$ is the Heaviside function. $f_0^{l}$ and $f_0^{r}$ are the initial gas distribution functions on the left and right sides of the interface, which are determined by corresponding initial macroscopic variables and their spatial derivatives.
  With the second-order accuracy, $f_0^k(\boldsymbol{x}_{0},\boldsymbol{v}_0)$ is constructed by Taylor expansion around $(\boldsymbol{x},\boldsymbol{v})$
  \begin{equation}\label{taylor_ex}
    f_0^k(\boldsymbol{x}_{0},\boldsymbol{v}_0)=f_G^k (\boldsymbol{x},\boldsymbol{v}) - \boldsymbol{w}t\cdot \nabla_xf_G^k +(\boldsymbol{\Omega}\times \boldsymbol{v})t \cdot \nabla_{v} f_G^k
  \end{equation}
 for $k=l,r$. Due to Chapman-Enskog expansion, $f_G^k$ is given by 
  \begin{equation}\label{ce_ex}
    f_G^k= g^k[1-\tau(A^k +  \boldsymbol{a}^k \cdot \boldsymbol{w})]+ \tau (\boldsymbol{\Omega}\times \boldsymbol{v}) \cdot \nabla_{v} g^k,
  \end{equation}
where $g^k$ is the equilibrium distribution function defined by the macroscopic variables $\boldsymbol{W}^k$ at the both sides of a cell interface, $\boldsymbol{a}^k$ are defined by the spatial derivatives of $g^k$
\begin{equation*}
	\begin{aligned}
a_i^k&= (\frac{\partial g^k}{\partial \rho^k }\frac{\partial \rho^k}{\partial x_i}+ \frac{\partial g^k}{\partial V_1^k }\frac{\partial V_1^k}{\partial x_i} + \frac{\partial g^k}{\partial V_2^k }\frac{\partial V_2^k}{\partial x_i} + \frac{\partial g^k}{\partial V_3^k }\frac{\partial V_3^k}{\partial x_i}+ \frac{\partial g^k}{\partial \lambda^k}\frac{\partial \lambda^k}{\partial x_i})/g^k
\\&= a_{i1}^k +a_{i2}^k v_1 + a_{i3}^k v_2 + a_{i4}^k v_3 +a_{i5}^k \frac{1}{2}(v_1^2+v_2^2+v_3^2\xi^2),
	\end{aligned}
\end{equation*}
and
  \begin{equation*}
A^k=A_{1}^k +A_{2}^k v_1 + A_{3}^k v_2 + A_{4}^k v_3 +A_{5}^k \frac{1}{2}(v_1^2+v_2^2+v_3^2\xi^2)
\end{equation*}
are determined  by compatibility condition
  \begin{equation}\label{compatibility}
\int(f^k_G-g^k) \Psi\td\boldsymbol{v}\td\Xi=0.
\end{equation}
  Substituting Eq.(\ref{taylor_ex}) and (\ref{ce_ex}) into (\ref{ini_gas}), the initial gas distribution has following form
  \begin{equation}\label{inidis}
    \begin{aligned}
    f_{0}= \begin{cases}g^{l}\left[1-\left(\boldsymbol{a}^{l} \cdot \boldsymbol{w}\right) t-\tau\left(A^{l}+\boldsymbol{a}^{l} \cdot \boldsymbol{w} \right)\right]+(t+\tau)\left(\boldsymbol{\Omega}\times \boldsymbol{v})  \cdot \nabla_{v} g^{l}\right), & x_1<0 , \\
      g^{r}\left[1-\left(\boldsymbol{a}^{r} \cdot \boldsymbol{w}\right) t-\tau\left(A^{r}+\boldsymbol{a}^{r} \cdot \boldsymbol{w}\right)\right]+(t+\tau)\left(\boldsymbol{\Omega}\times \boldsymbol{v}) \cdot \nabla_{v} g^{r}\right), & x_1 \geq 0 ,\end{cases}
    \end{aligned}
    \end{equation}
Then, the equilibrium distribution is defined by the Taylor expansion
    \begin{equation}\label{equdis}
      \begin{aligned}
        g(\boldsymbol{x^{\prime}},t^{\prime},\boldsymbol{v^{\prime}})=&\overline{g}(\boldsymbol{x},0,\boldsymbol{v}) + \nabla_x \overline{g}\cdot (\boldsymbol{x^{\prime}} -\boldsymbol{x}) + \nabla_v \overline{g}\cdot (\boldsymbol{v^{\prime}} -\boldsymbol{v}) + \overline{g}_t t^{\prime} \\
        =&\overline{g}(\boldsymbol{x},0,\boldsymbol{v}) - \nabla_x \overline{g}\cdot \boldsymbol{w}(t-t^{\prime})  + \nabla_v \overline{g}\cdot \left(\boldsymbol{\Omega}\times\boldsymbol{v}\right) (t-t^{\prime}) + \overline{g}_t t^{\prime} \\
        =&\overline{g}(\boldsymbol{x},0,\boldsymbol{v}) \left\{ 1 - \overline{\boldsymbol{a}}\cdot \boldsymbol{w}(t-t^{\prime})+A t^{\prime} \right\} + \nabla_v \overline{g}\cdot \left(\boldsymbol{\Omega}\times\boldsymbol{v}\right) (t-t^{\prime}), \\
       \end{aligned}
    \end{equation}
where $\overline{g}$ and $\overline{\boldsymbol{a}}$ are determined from the reconstruction of macroscopic flow variables presented in section \ref{equ_res}, and $A$ is obtained by compatibility condition \eqref{compatibility}.
By substituting  Eq. (\ref{inidis}) and Eq. (\ref{equdis}) into Eq. (\ref{solbgk}) and keeping the second order accuracy,
the solution at a cell interface becomes
  \begin{equation}\label{solution}
    \begin{aligned}
    f\left(\boldsymbol{x}, t, \boldsymbol{v}, \xi\right)
    &=(1-e^{-t / \tau_n})\bar{g}+e^{-t / \tau_n}\left[\mathbb{H}\left(w_1\right) g^{l}+\left(1-\mathbb{H}\left(w_1\right)\right) g^{r}\right] +t\bar{A} \bar{g}\\
    &-\tau\left(1-e^{-t / \tau_n}\right)\left[ \left(\overline{\boldsymbol{a}} \cdot \boldsymbol{w}\right)\bar{g}-(\boldsymbol{\Omega}\times \boldsymbol{v}) \cdot \nabla_{v} \bar{g}+\bar{A} \bar{g}\right]\\
    &-\tau e^{-t / \tau_n} \mathbb{H}\left(w_1\right)\left[\left(\boldsymbol{a}^{l} \cdot \boldsymbol{w}\right) g^{l}-(\boldsymbol{\Omega}\times \boldsymbol{v}) \cdot \nabla_{v} g^{l}+ A^{l}g^{l}\right]\\
    &-\tau e^{-t / \tau_n} \left(1-\mathbb{H}\left(w_1\right)\right)\left[\left(\boldsymbol{a}^{r} \cdot \boldsymbol{w}\right) g^{r}-(\boldsymbol{\Omega}\times \boldsymbol{v}) \cdot \nabla_{v} g^{r} + A^{r}g^{r} \right]\\
    &+t e^{-t / \tau_n}\left[ \left(\overline{\boldsymbol{a}} \cdot \boldsymbol{w}\right)\bar{g}- \mathbb{H}\left(w_1\right)\left(\boldsymbol{a}^{l} \cdot \boldsymbol{w}\right) g^{l}-\left(1-\mathbb{H}\left(w_1\right)\right)\left(\boldsymbol{a}^{r} \cdot \boldsymbol{w}\right) g^{r}\right].\\
  \end{aligned}
    \end{equation}
The fluxes in Eq.(\ref{flux}) can be obtained by taking the moments of the above distribution function.
The calculation of moments can be found in \ref{moments}.

\subsection{Evolution of the cell-averaged spatial gradients}
 By taking moments of the above gas distribution function in Eq. (\ref{solution}), the time-accurate conservative flow variables
 at a cell interface can be also obtained
\begin{equation}\label{conval}
  \boldsymbol{W}_{p,k}(t^{n+1}) = \mathbf{T}^\prime\left(\int  \widetilde{\boldsymbol{ \Psi}} f(\tilde{\boldsymbol{x}}_{p,k},t^{n+1},\tilde{\boldsymbol{v}},\xi) \td\boldsymbol{v}\td\Xi \right).
\end{equation}
According to Divergence theorem, the cell averaged gradients over cell $\Omega_i$ at time $t^{n+1}$ are
\begin{equation}\label{slop1}
  \overline{\nabla \boldsymbol{W}}^{n+1}_i = \frac{1}{|\Omega_i|}\sum_{p=1}^{N_f}\int_{\Gamma_{ip}} \boldsymbol{W}^{n+1} \boldsymbol{n}_p dS,
\end{equation}
where the surface integration can be calculated by Gaussian quadrature (the index $i$ is omitted )
\begin{equation}\label{slop2}
  \int_{\Gamma_{p}} \boldsymbol{W}^{n+1}  \boldsymbol{n}_p dS\approx \sum_{k=1}^{M_p}|S_p| \omega_k \boldsymbol{W}_{p,k}(t^{n+1})\boldsymbol{n}_{p,k}.
\end{equation}
Besides evaluating the cell averaged gradients,  the solution updates of the scheme are presented next.

\section{Solution updates and temporal discretization}
According to the semi-discretization Eq. (\ref{fvm}), the right side contains two parts, the net flux $\mathcal{L}_F$ and the source term.
The two-stage fourth-order (S2O4) temporal discretization is adopted here for the solution updates \cite{liTwoStageFourthOrder2016},
\begin{equation*}
  \begin{aligned}
    \boldsymbol{W}_{i}^{*\prime}=& \boldsymbol{W}_{i}^{n}+\frac{1}{2} \Delta t \mathcal{L}_F\left(\boldsymbol{W}_{i}^{n}\right)+\frac{1}{8} \Delta t^{2} \frac{\partial}{\partial t} \mathcal{L}_F\left(\boldsymbol{W}_{i}^{n}\right) ,\\
    \boldsymbol{W}_{i}^{*}=& \boldsymbol{W}_{i}^{*\prime}+\int_{t_n}^{t_n+\frac{1}{2}\Delta t} \boldsymbol{S}(\boldsymbol{W}_i^{*\prime})\td t,\\
    \boldsymbol{W}_{i}^{(n+1)\prime}=& \boldsymbol{W}_{i}^{n}+\Delta t \mathcal{L}_F\left(\boldsymbol{W}_{i}^{n}\right) +\frac{1}{6} \Delta t^{2}\left(\frac{\partial}{\partial t} \mathcal{L}_F\left(\boldsymbol{W}_{i}^{n}\right)+2 \frac{\partial}{\partial t} \mathcal{L}_F\left(\boldsymbol{W}_{i}^{*}\right)\right)\\
    \boldsymbol{W}_{i}^{n+1}=& \boldsymbol{W}_{i}^{(n+1)\prime}+\int_{t_n}^{t_n+\Delta t} \boldsymbol{S}(\boldsymbol{W}_i^{(n+1)\prime})\td t.
  \end{aligned}
  \end{equation*}
The source term only appears in moment equations ($\frac{\td \rho \boldsymbol{V}}{\td t} =-\boldsymbol{\Omega}\times (\rho \boldsymbol{V}) $),
which is integrated as 
  \begin{equation*}
    \begin{aligned}
    \int_{t_n}^{t_n+\Delta t} -\boldsymbol{\Omega}\times (\rho \boldsymbol{V}) \td t =&-\left(\rho_n \boldsymbol{V}_n \times  \frac{\boldsymbol{\Omega}}{\Omega}\right)\sin{(\Omega \Delta t)}\\
    &- \left[\rho_n \boldsymbol{V}_n -\left(\rho_n \boldsymbol{V}_n \cdot \frac{\boldsymbol{\Omega}}{\Omega}\right) \frac{\boldsymbol{\Omega}}{\Omega}\right](1-\cos{(\Omega \Delta t)}).
    \end{aligned}
  \end{equation*}
The time-dependent gas distribution function at Gauss points on the interfaces is updated by
  \begin{equation*}
    \begin{aligned}
    f^{*} &=f^{n}+\frac{1}{2} \Delta t f_{t}^{n}, \\
    f^{n+1} &=f^{n}+\Delta t f_{t}^{*},
    \end{aligned}
    \end{equation*}
where the time-dependent conservative values at each Gauss point can be obtained by Eq. (\ref{conval}). Then by Eq. (\ref{slop1}) and Eq. (\ref{slop2}), the cell-averaged slopes can be updated.

\section{HWENO Reconstruction}
The 3rd-order compact reconstruction \cite{jiHWENOReconstructionBased2020}  is adopted here with cell-averaged values and cell-averaged first-order spatial derivative. In order to capture shock, WENO weights \cite{zhu2020} and gradient compression factor (CF) \cite{jiGC2021} are used. In this work, we further improve the WENO procedures and CF with the consideration of simplicity and robustness. Only one large stencil and one sub stencil are involved in the new WENO procedure.
\subsection{3rd-order compact reconstruction for large stencil}
Firstly, a linear reconstruction is presented. To achieve a third-order accuracy in space, a quadratic polynomial $p^2$ is constructed as follows
\begin{equation*}
  \begin{aligned}
    p^2&=a_0+\frac{1}{h}[a_1(x-x_0)+a_2(y-y_0)+a_3(z-z_0)]\\
    &+\frac{1}{2h^2}[a_4(x-x_0)^2+a_5(y-y_0)^2+a_6(z-z_0)^2]\\
    &+\frac{1}{h^2}[a_7(x-x_0)(y-y_0)+a_8(y-y_0)(z-z_0)+a_9(x-x_0)(z-z_0)],
  \end{aligned}
\end{equation*}
where $h=\frac{V}{\max_j S_j}$($V$ is the volume of cell and $S_j$ is the area of cell's surface ) is the cell size, and $(x_0,y_0,z_0)$ is the coordinate of cell center.

The $p^2$ on $\Omega_0$  is constructed on the compact stencil $S$ including $\Omega_0$ and its all von Neumann neighbors $\Omega_m$($m=1,\cdots,N_f$, where $N_f=6$ for hexahedron cell or $N_f=5$  for triangular prism). The cell averages $\overline{Q}$ over $\Omega_0$ and $\Omega_m$ and cell averages of space partial derivatives $\overline{Q}_x,\overline{Q}_y $ and $\overline{Q}_z$ over $\Omega_m$ are used to obtain $p^2$.

The polynomial $p^2$ is required to exactly satisfy cell averages over both $\Omega_0$ and $\Omega_m$ ($m=1,\cdots,N_f$)
\begin{equation*}
  \iiint_{\Omega_0} p^2 \text{d}V = \overline{Q}_0|\Omega_0|,\iiint_{\Omega_m} p^2 \text{d}V = \overline{Q}_m|\Omega_m| ,
\end{equation*}
with the following condition satisfied in a least-square sense
\begin{equation*}
  \begin{aligned}
    \iiint_{\Omega_m}\frac{\partial}{\partial x} p^2 \text{d}V = \left(\overline{Q}_x\right)_m|\Omega_m|\\
    \iiint_{\Omega_m}\frac{\partial}{\partial y} p^2 \text{d}V = \left(\overline{Q}_y\right)_m|\Omega_m|\\
    \iiint_{\Omega_m}\frac{\partial}{\partial z} p^2 \text{d}V = \left(\overline{Q}_z\right)_m|\Omega_m|.\\
  \end{aligned}
\end{equation*}
To solve the above system, the constrained least-square method is used.
\subsection{Green-Gauss reconstruction for the sub stencil}
The classical Green-Gauss reconstruction with only cell-averaged values is adopted to provide the linear polynomial $p^1$ for the sub stencil.
\begin{equation*}
  p^1=\overline{Q}+\boldsymbol{x}\cdot \sum_{m=1}^{N_f}\frac{\overline{Q}_m+\overline{Q}_0}{2}S_m\boldsymbol{n} _m,
\end{equation*}
where $S_m$ is the area of the cell's surface and $\boldsymbol{n}_m$ is the surface's normal vector.
\subsection{Gradient compression Factor}
The CF was first proposed in \cite{jiGC2021}. Here several improvements have been made: there is no $\epsilon$ in the improved expression of CF; the difference of Mach number is added for improving the robustness under strong rarefaction wave. Denote $\alpha_i \in[0,1]$ as gradient compression factor at targeted cell $\Omega_i$
\begin{equation*}
  \alpha_i = \prod_{p=1}^m\prod_{k=0}^{M_p}\alpha_{p,k},
\end{equation*}
where $\alpha_{p,k}$ is the CF obtained by the $k$th Gaussian point at the interface $p$ around cell $\Omega_i$, which can be calculated by
\begin{equation*}
  \begin{aligned}
    &\alpha_{p,k}=\frac{1}{1+A^2}, \\
    &A=\frac{|p^l-p^r|}{p^l} +\frac{|p^l-p^r|}{p^r}+(\text{Ma}^{l}_n-\text{Ma}^{r}_n)^2+(\text{Ma}^{l}_t-\text{Ma}^{r}_t)^2,
  \end{aligned}
\end{equation*}
where $p$ is pressure, $\text{Ma}_n$ and $\text{Ma}_t$ are the Mach numbers defined by normal and tangential velocity, and superscript $l,r$ denote the left and right values of the Gaussian points.

Then, the updated slope is modified by
\begin{equation*}
  \widetilde{\overline{\nabla \boldsymbol{W}}}_i^{n+1} = \alpha_i\overline{\nabla \boldsymbol{W}}_i^{n+1},
\end{equation*}
and the Green-Gauss reconstruction is modified as
\begin{equation*}
  p^1=\overline{Q}+\alpha \boldsymbol{x}\cdot \sum_{m=1}^{N_f}\frac{\overline{Q}_m+\overline{Q}_0}{2}S_m\boldsymbol{n} _m.
\end{equation*}
\subsection{Non-linear WENO weights}
In order to deal with discontinuity, the idea of multi-resolution WENO reconstruction is adopted \cite{jiGC2021,zhu2020}. 
Here only two polynomials are chosen
\begin{equation*}
  P_2=\frac{1}{\gamma_2}p^2-\frac{\gamma_1}{\gamma_2}p^1  ,P_1 =p^1.
\end{equation*}
Here, we choose $\gamma_1=\gamma_2=0.5$. So the quadratic polynomial $p^2$ can be written as
\begin{equation}\label{weno-linear}
  p^2 = \gamma_1P_1 + \gamma_2 P_2.
\end{equation}
Then, we can define the smoothness indicators
\begin{equation*}
  \beta_{j}=\sum_{|\alpha|=1}^{r_{j}}\Omega^{\frac{2}{3}|\alpha|-1} \iiint_{\Omega}\left(D^{\alpha} p^{j}(\mathbf{x})\right)^{2} \mathrm{~d} V,
  \end{equation*}
where $\alpha$ is a multi-index and $D$ is the derivative operator, $r_1=1,r_2=2$. Special care is given for $\beta_1$ for better robustness
  \begin{equation*}
    \beta_1=\min(\beta_{1,\text{Green-Gauss}},\beta_{1,\text{least-square}}),
  \end{equation*}
where $\beta_{1,\text{Green-Gauss}}$ is the smoothness indicator defined by Green-Gauss reconstruction, and $\beta_{1,\text{least-square}}$ is the smoothness indicator defined by second-order least-square reconstruction. Then, the smoothness indicators $\beta_i$ are non-dimensionalized by
  \begin{equation*}
    \tilde{\beta}_i=\frac{\beta_i}{Q_0^2+\beta_1+10^{-40}}.
  \end{equation*}
The nondimensionalized global smoothness indicator $\tilde{\sigma}$ can be defined as
  \begin{equation*}
    \tilde{\sigma}=\left|\tilde{\beta}_{1}-\tilde{\beta}_{0}\right|^{\frac{4}{3}}.
    \end{equation*}
Therefore, the corresponding non-linear weights are given by
    \begin{equation*}
      \tilde{\omega}_{m}=\gamma_{m}\left(1+\left(\frac{\tilde{\sigma}}{\epsilon+\tilde{\beta}_{m}}\right)^{2}\right),\epsilon=10^{-5},
      \end{equation*}
      \begin{equation*}
        \bar{\omega}_{m}=\frac{\tilde{\omega}_{m}}{\sum \tilde{\omega}_{m}}, m=1,2.
      \end{equation*}
Replacing $\gamma_m$ in equation (\ref{weno-linear}) by $\bar{\omega}_{m}$ , the final non-linear reconstruction can be obtained
      \begin{equation*}
        R(\boldsymbol{x})=\bar{\omega}_2P_2+\bar{\omega}_1 P_1.
      \end{equation*}
The desired non-equilibrium states at Gaussian points become
\begin{equation*}
  Q_{p, k}^{l, r}=R^{l, r}\left(\boldsymbol{x}_{p, k}\right), \left(Q_{x_{i}}^{l, r}\right)_{p, k}=\frac{\partial R^{l, r}}{\partial x_{i}}\left(\boldsymbol{x}_{p, k}\right).
  \end{equation*}
\subsection{Reconstruction of equilibrium states}\label{equ_res}
After reconstructing the non-equilibrium state, a kinetic weighted average method can be used to get equilibrium states 
and tangential derivatives \cite{jiHWENOReconstructionBased2020},
\begin{equation*}
  \int \bar{g}\Psi\td\boldsymbol{v}\td\Xi =\boldsymbol{W}_0 = \int_{\tilde{w}_1>0} g^{l} \Psi\td\boldsymbol{v}\td\Xi +\int_{\tilde{w}_1<0} g^{r}\Psi\td\boldsymbol{v}\td\Xi,
\end{equation*}
\begin{equation*}
  \int \bar{a}_i \bar{g}\Psi\td\boldsymbol{v}\td\Xi = \frac{\partial \boldsymbol{W}_{0} }{\partial \tilde{x}_i}= \int_{\tilde{w}_1>0}a_i^l g^{l} \Psi\td\boldsymbol{v}\td\Xi +\int_{\tilde{w}_1<0} a_i^r g^{r}  \Psi\td\boldsymbol{v}\td\Xi  , i=2,3.
\end{equation*}
For the normal derivatives, the above solution is further modified according to the idea in linear diffusive generalized Riemann problem (dGRP) \cite{dgrp}
\begin{equation}\label{dgrp}
  \int \bar{a}_1\bar{g}\Psi\td\boldsymbol{v}\td\Xi = \frac{\partial \boldsymbol{W}_{0} }{\partial \tilde{x}_1}=  \int_{\tilde{w}_1>0} a_1^l g^{l}  \Psi\td\boldsymbol{v}\td\Xi +\int_{\tilde{w}_1<0} a_1^r g^{r}  \Psi\td\boldsymbol{v}\td\Xi + \frac{\boldsymbol{W}^r-\boldsymbol{W}^l}{(\boldsymbol{x}_{rc}-\boldsymbol{x}_{lc}) \cdot \boldsymbol{n}},
\end{equation}
where $\boldsymbol{x}_{rc}$ and $\boldsymbol{x}_{lc}$ are the coordinates of left and right cell centroid, and $\boldsymbol{n}$ is the normal vector of interface. By adding a penalty term in Eq. (\ref{dgrp}), the whole scheme is essentially free from the odd-even decoupling phenomenon \cite{BLAZEK201573}.
\section{Sliding mesh method}
To simulate the problem with sliding interface, the computational domain is divided into rotating and stationary parts.
The whole computational algorithm is shown as Algorithm \ref{alg1}, where the bold text is special treatments relating to the sliding interface. The detailed algorithm will be discussed in the following subsections.
\begin{algorithm}
	%\textsl{}\setstretch{1.8}
	\renewcommand{\algorithmicrequire}{\textbf{Input:}}
	\renewcommand{\algorithmicensure}{\textbf{Output:}}
  \caption{CGKS with sliding interface}
  \label{alg1}
  \begin{algorithmic}[1]
    \STATE Label the interface between rotor and stator, and the neighbor cells of the interface
    \STATE Initial rotation angle $\alpha=0$
    \WHILE{the computation uncompleted}
    \STATE calculate the time step $\Delta t$ according to CFL number
      \STATE {\bf{ rotate the interface $\alpha$, establish mortars}}
      \FOR{i=1,2(for S2O4)}
      \STATE define boundary condition for ghost cell
      \STATE reconstruct the cell distribution except the adjoint cells of the sliding interface
      \STATE {\bf{reconstruct the adjoint cells of sliding interface}}
      \STATE define boundary condition at boundary Gaussian point
      \STATE evolution for interfaces
      \STATE {\bf{evolution for mortars}}
      \STATE update cell average conservative values
      \STATE add source term for the rotating domain
      \STATE update cell average first-order spatial derivatives of conservative values except the adjoint cells of the sliding interface
      \STATE {\bf{ update cell average first-order spatial derivatives of conservative values for the adjoint cells of the sliding interface} }
      \ENDFOR
      \STATE $\alpha = \alpha + \omega *\Delta t$
    \ENDWHILE
  \end{algorithmic}
  \end{algorithm}
\subsection{Establish mortar by polygon clipping}
\begin{figure}[hbt!]
  \centering
  \subfigure[The intersecting of two meshes in 2-D\label{sli_ex}]{\includegraphics[width=6cm]{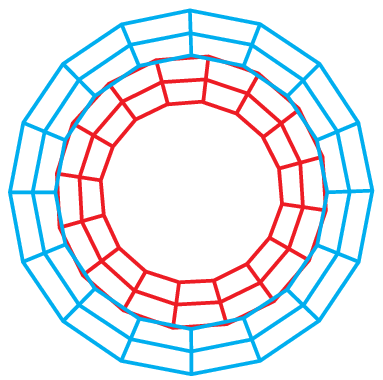}}
  \subfigure[The mortars in 2-D \label{sli_ex2}]{\includegraphics[width=6cm]{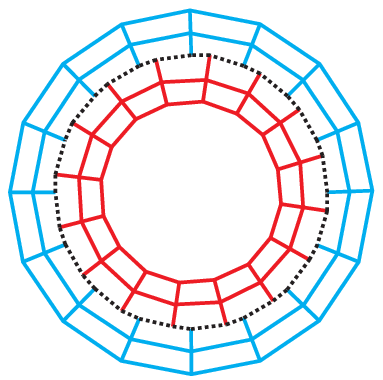}}
  \subfigure[Clipping result\label{clip_res}]{\includegraphics[width=6cm]{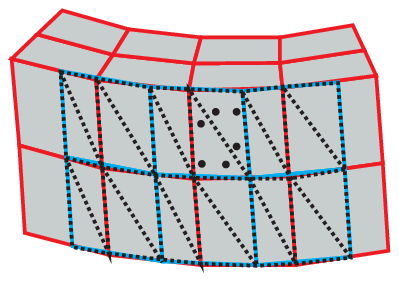}}
  \caption[]{Polygon clipping process}
\end{figure}

To communicate the information between the rotating part and the stationary part, the mortar elements need to be established. As shown in Fig. \ref{sli_ex}, the rotating and stationary parts overlap on the same circle but do not have common interfaces. So, new mortar interfaces, as shown by black dotted lines in Fig. \ref{sli_ex2}, need to be defined.

The in-house 3-D code based on prism mesh is used for the current simulation. The interfaces between the rotating part and the stationary part are surface meshes. The Sutherland-Hodgman algorithm \cite{Sutherland197432} is used for clipping the intersecting polygon of the two connected elements on the interface. This method is an effective and accurate algorithm for convex polygon clipping, which could deal with triangles, quadrangles, and so on. We triangulate the clipped polygon to make the algorithm easily adapt to different intersecting polygons.

As shown in Fig. \ref{sli_ex}, the sliding interface is a cylindrical surface, but straight edge meshes are used in the computation.
We transform this interface into the cylindrical coordinate and consider that all the nodes have the same radius, so the polygon clipping process is done in $\theta-z$ plane. As shown in Fig. \ref{clip_res}, the black dotted lines show generated mortars by two interfaces, where the red line shows the inner part and blue line the outer part.
\subsection{Reconstruction for sliding mesh}
\begin{figure}[hbt!]
  \centering
   \includegraphics[width=2cm]{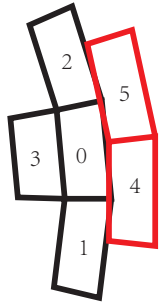}
  \caption[]{reconstruction stencil for sliding mesh}\label{sliding_stencil}
\end{figure}

\begin{figure}[hbt!]
  \centering
  \subfigure[]{\includegraphics[width=4cm]{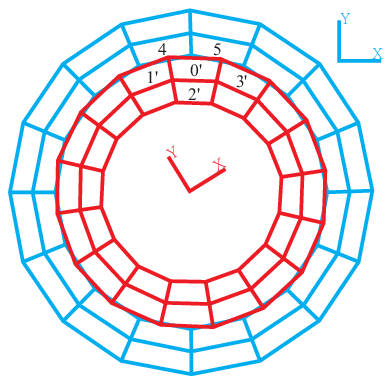}}
  \subfigure[]{\includegraphics[width=4cm]{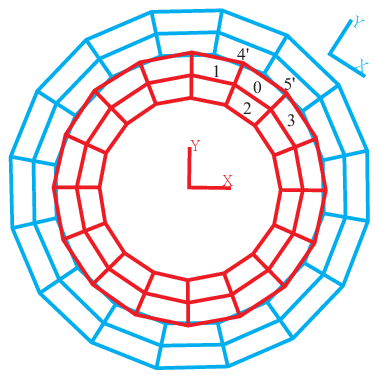}}
  \caption[]{coordinate transformation}\label{stencil}
\end{figure}

To reconstruct the adjoint cells of sliding interface, the same stencils are used for these cells. However, one of the interfaces is the sliding interface, so there is no directly jointing neighbor cell at this interface. As shown in Fig. \ref{sliding_stencil},
the right face of cell 0 is a sliding interface, where two cells (cell 4 and cell 5) joint with cell 0. 
Under this condition, a ghost cell is created by merging cell 4 and cell 5. For third-order reconstruction, the new constraints become
\begin{equation*}
  \begin{aligned}
    \iiint_{\Omega_4+\Omega_5}p^2 \text{d}V &= \overline{Q}_4|\Omega_4|+\overline{Q}_5|\Omega_5|,\\
    \iiint_{\Omega_4+\Omega_5}\frac{\partial}{\partial x} p^2 \text{d}V &= \left(\overline{Q}_x\right)_4|\Omega_4|+\left(\overline{Q}_x\right)_5|\Omega_5|,\\
    \iiint_{\Omega_4+\Omega_5}\frac{\partial}{\partial y} p^2 \text{d}V &= \left(\overline{Q}_y\right)_4|\Omega_4|+\left(\overline{Q}_y\right)_5|\Omega_5|,\\
    \iiint_{\Omega_4+\Omega_5}\frac{\partial}{\partial z} p^2 \text{d}V &= \left(\overline{Q}_z\right)_4|\Omega_4|+\left(\overline{Q}_z\right)_5|\Omega_5|.
  \end{aligned}
\end{equation*}
And for second-order Green-Gauss reconstruction, the cell averaged conservative flow variables of the ghost cell are obtained by the volume weighted averages of the  values in cell 4 and cell 5. In addition, special treatment should be taken due to coordinate transformation. As shown in Fig. \ref{stencil},
the rotor has rotated by a certain degree, and the cells 0, 1, 2, and 3 have moved to the position of $0^{'}$, $1^{'}$, $2^{'}$, $3^{'}$. However, the governing equations in the related framework are used to simulate the rotation effect. So the cells are still in the position of 0, 1, 2, 3. To reconstruct cell 0, we need to rotate cells 4 and 5 to the position $4^{'}$ and $5^{'}$ with transformation of both geometry information and cell-average values, including conservation values and their spatial derivatives.
\subsection{Flux evaluation via mortars and cell-average slope update}
After the mortars are created, the origin interface will be replaced by mortars. So we need to set Gaussian points (as shown by black dots
in Fig. \ref{clip_res}) in the mortars, and the left and right values of Gaussian points can be obtained by reconstruction. After reconstruction, the fluxes via mortar and point-wise conservative variables of Gaussian points on mortars can be updated.

\section{Numerical experiments}
In the following cases, the three-dimensional solver is used to solve two-dimensional problems. 
Two layers and periodic boundary conditions are used in the $z$ direction. The time step is given by $\Delta t= \min{\Delta t_i}$, where $\Delta t_i$ is the time step defined in each cell
\begin{equation*}
  \Delta t_i=C_{\text{CFL}}\frac{h_i}{|\boldsymbol{V}-\boldsymbol{U}|_i+c_i+2\nu_i/h_i},
\end{equation*}
where $C_{\text{CFL}}$ is the CFL number, $|\boldsymbol{V}-\boldsymbol{U}|_i$, $c_i$ , and $\nu_i=(\mu/\rho)_i$ are the magnitude of related velocities, sound speed and kinematic viscosity coefficient of cell $i$. Here, we set the CFL number as 0.5.
\subsection{Isentropic vortex propagation}
The isentropic vortex propagation problem is selected to test the solver for inviscid flow.
The computation domain is $[0,1]\times[0,1]$. The flow at time $t$ is
\begin{equation*}
  \begin{aligned}
  &(U, V)=(U_0,V_0)+\frac{\kappa}{2 \pi} e^{0.5\left(1-r^{2}\right)}(-\bar{y}, \bar{x}), \\
  &T=1-\frac{(\gamma-1) \kappa^{2}}{8 \gamma \pi^{2}} e^{1-r^{2}} \\
  &S=1,\\
  &T=\frac{p}{\rho} , S=\frac{p}{\rho^\gamma},
  \end{aligned}
  \end{equation*}
  where the non-dimensional coordinate is $(\bar{x},\bar{y})=(\frac{x-x_r}{r_0},\frac{y-x_r}{r_0}),r_0=0.05$, the radius 
  $r =\sqrt{\bar{x}^2+\bar{y}^2}$, and the vortex strength $\kappa=5$. The $(x_r,y_r)$ depends on time
  \begin{equation*}
    x_r=x_0-U_0t,y_r=y_0-V_0t.
  \end{equation*}
  In our simulation, we choose $(x_0,y_0)=(0.5,0.5)$ and the background velocity $(U_0,V_0)=(1,1)$. Periodic boundary conditions are applied in both the x and y directions.

\begin{figure}[htb!]
  \centering
  \includegraphics[width=6cm]{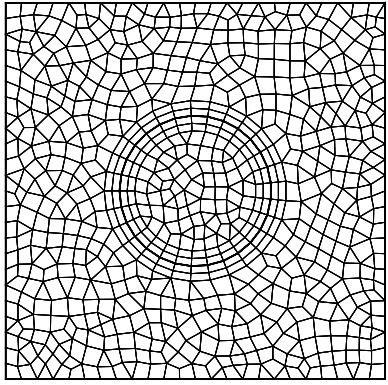}
  \caption{The coarsest mesh used for isentropic vortex propagation(the inner black circle indicates sliding interface) }\label{entro_mesh}
\end{figure}

The computation domain is divided into two parts: the rotating inner part with a radius of 0.2 and the stationary outer part.
The angle speed of the rotating part is set as $\omega=2\pi$. Both the rotating and stationary cases are calculated to test our method. Four meshes with cell number $1754\times2, 7122\times2,29324\times2 ,120252\times 2$ are used. The coarsest mesh is shown in Fig. \ref{entro_mesh}.

To validate the accuracy of the scheme, the density error is defined at $t=1$, when the vortex has traveled for one period. The errors and numerical orders of the rotating and the stationary cases are shown in Table \ref{ent_er_r} and Table \ref{ent_er_n}. For both the stationary and the rotating cases, the numerical orders are close to the theoretical third-order accuracy; and the error of the rotating case is smaller than that of the stationary case due to the smaller time step used in the rotating case under the same CFL number.
  \begin{table}[htb!]
    \centering
    \caption{$\omega=2\pi \text{rad/s}$}\label{ent_er_r}
    \begin{tabular}{c|c|c|c|c|c|c}
      \hline
  mesh    &  $Error_{L^1}$& $O_{L^1}$  &  $Error_{L^2}$   &$O_{L^2}$       & $Error_{L^\infty}$&$O_{L^\infty}$  \\ \hline
 $1754\times2$  &  5.89E-04&		   & 7.93E-03	&	   & 4.00E-01	&    \\
 $7122\times2$  &  2.24E-04&	1.39&	2.91E-03	&1.45&	1.51E-01	&1.40\\
 $29324\times2$ &  4.43E-05&	2.34&	5.11E-04	&2.51&	2.27E-02	&2.73\\
 $120252\times2$&  7.12E-06&	2.64&	7.05E-05	&2.86&	3.96E-03	&2.52\\
\hline
    \end{tabular}
  \end{table}
  \begin{table}[htb!]
    \centering
    \caption{$\omega=0 $}\label{ent_er_n}
    \begin{tabular}{c|c|c|c|c|c|c}
      \hline
  mesh    &  $Error_{L^1}$& $O_{L^1}$  &  $Error_{L^2}$   &$O_{L^2}$       & $Error_{L^\infty}$&$O_{L^\infty}$  \\ \hline
  $1754\times2$&  5.87E-04&      & 7.95E-03&        &3.99E-01 &     \\
$7122\times2$&  2.27E-04& 1.37& 2.95E-03& 1.43  &1.51E-01 &1.41 \\
$29324\times2$&  4.56E-05&  2.32& 5.43E-04& 2.44  &2.26E-02 &2.74 \\
$120252\times 2$&  7.27E-06&  2.65& 8.58E-05& 2.66  &3.79E-03 &2.57 \\
  \hline
    \end{tabular}
  \end{table}

\begin{figure}[hbt!]
    \centering
    \includegraphics[width=6cm]{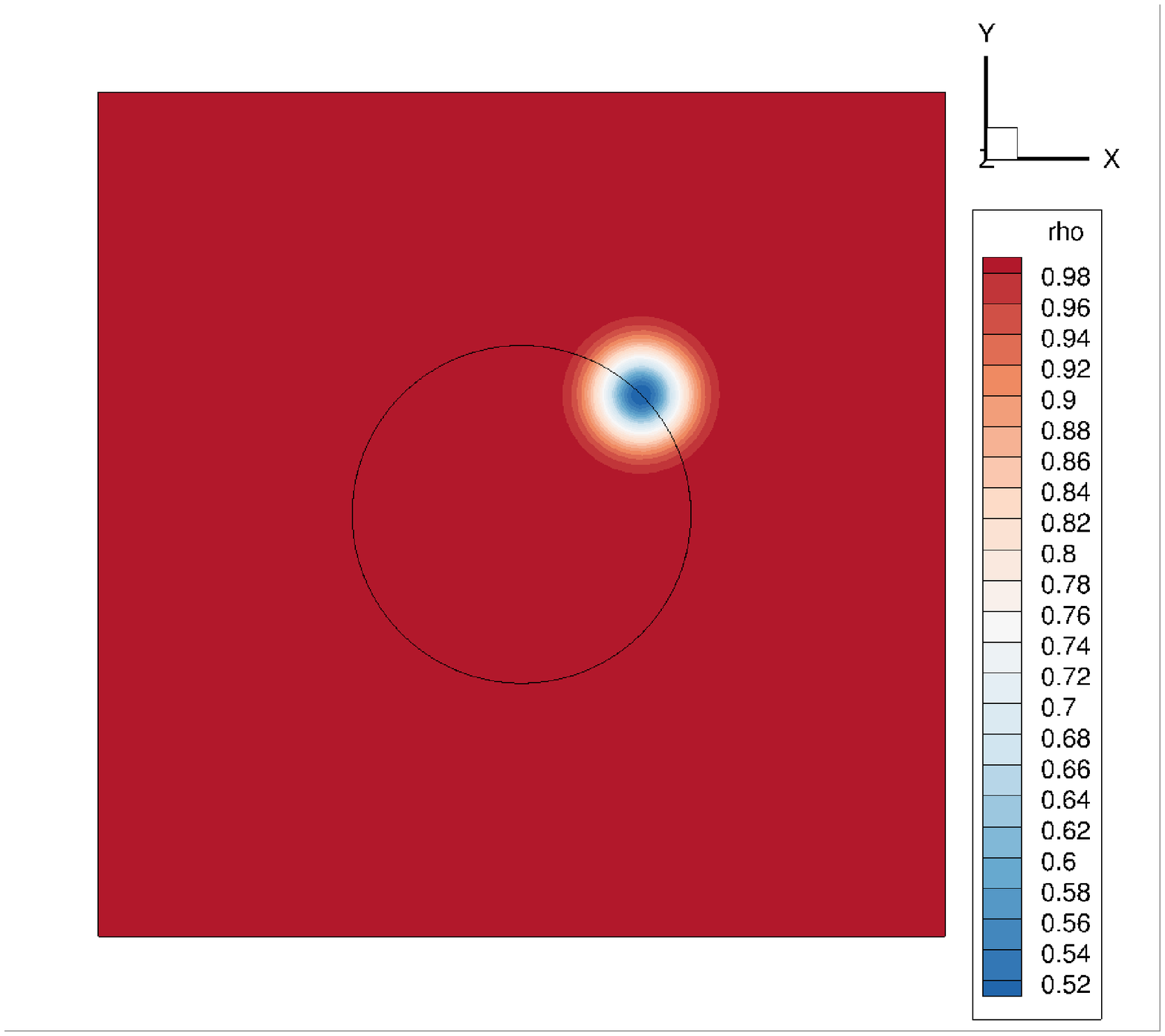}
    \includegraphics[width=6cm]{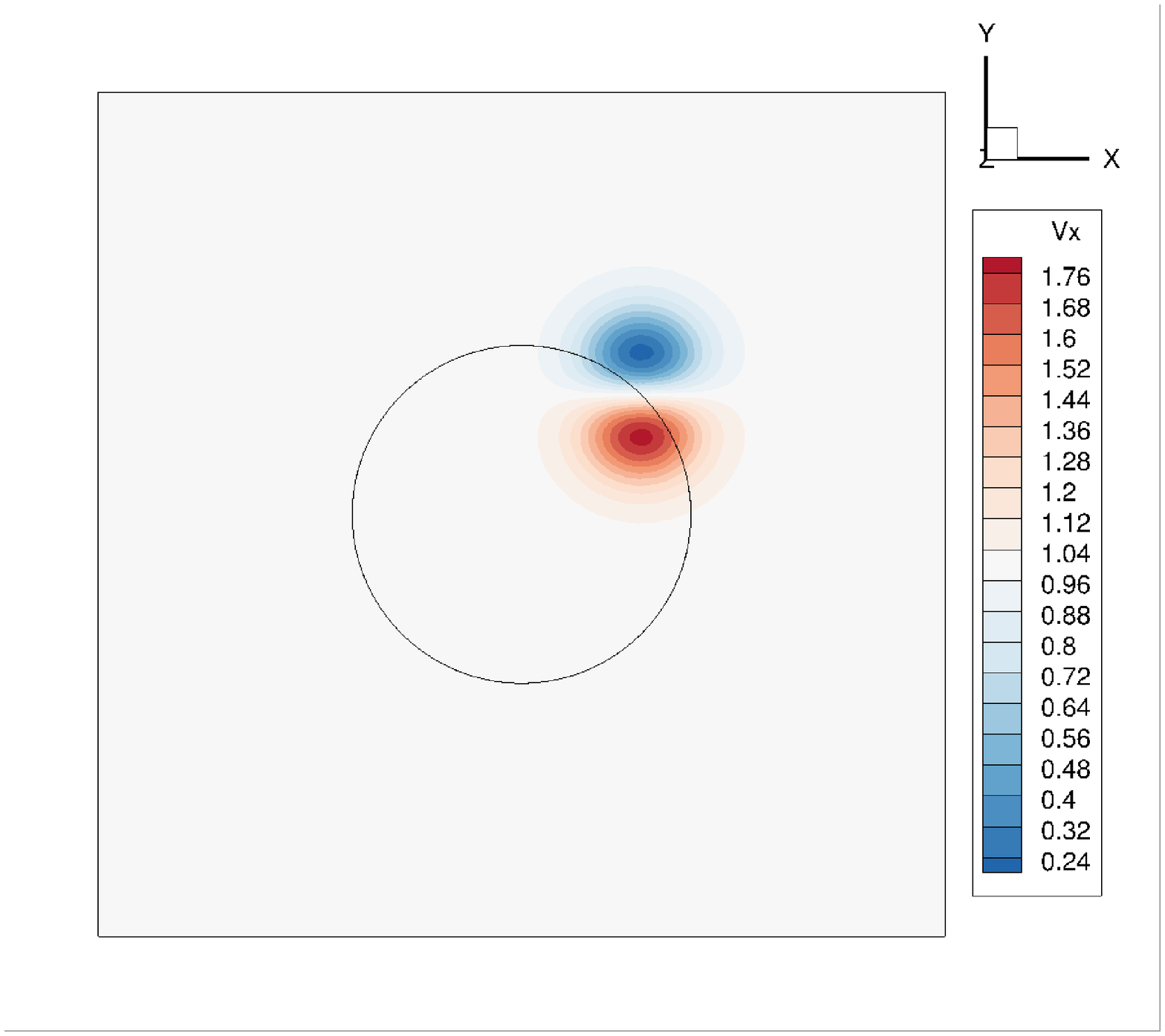}
    \caption{The density and velocity in $x$-direction contours of rotating case by finest mesh at $t=0.1\sqrt{2}$}\label{en_contour}
  \end{figure}

 The density and x-velocity contours of the rotating case by the finest mesh at $t=0.1\sqrt{2}$ are shown in Fig \ref{en_contour}, when the center of the vortex is located on the sliding interface. No distortion can be observed in the contour, which implies that the vortex can propagate through the sliding interface without reflection and deformation.

  \subsection{Flow over a rotating ellipse cylinder}
\begin{figure}[hbt!]
    \centering
    \includegraphics[width=6cm]{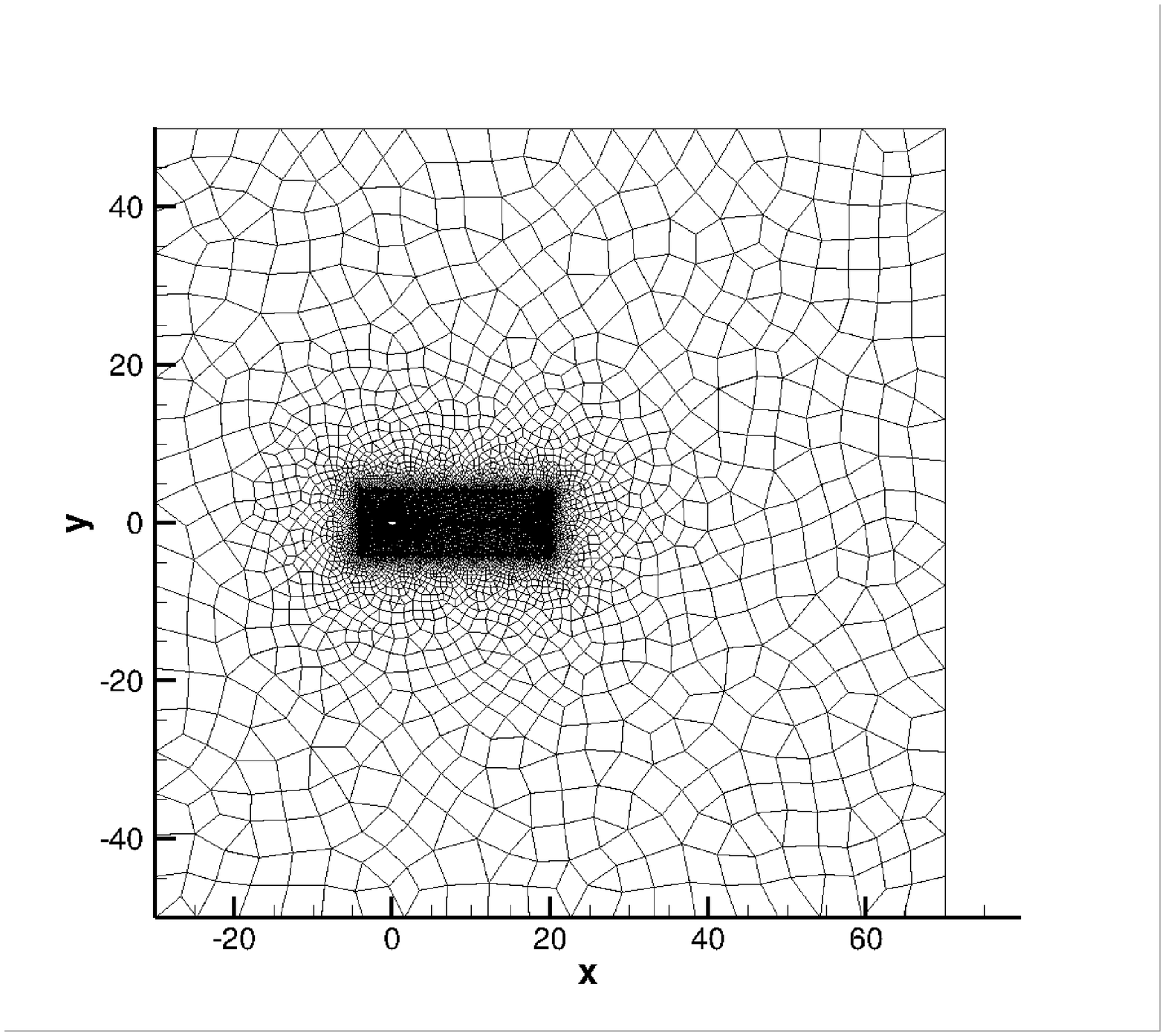}
    \includegraphics[width=6cm]{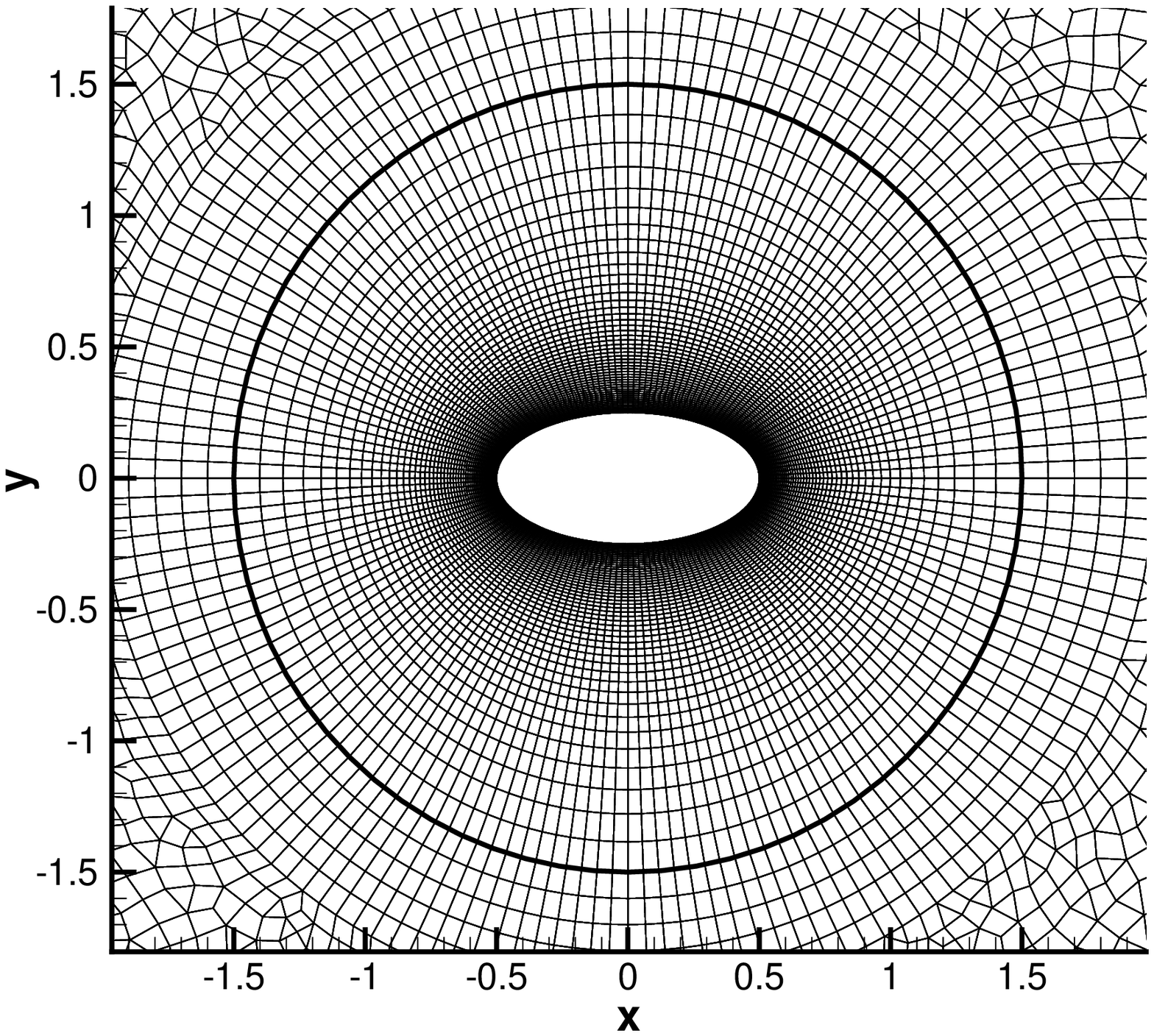}
    \caption{The mesh used in rotating ellipse cylinder case}\label{ell_mesh}
  \end{figure}

  This case is selected to verify our method for subsonic viscous flow. The income flow is set as $\rho_\infty=1.0, p_\infty=1/\gamma, U_\infty =0.05$, which has a Mach number 0.05. The ellipse, with a major axis length of $A=1.0$ and a minor axis length of $B=0.5$, rotates counterclockwise at an angular speed of $\omega=\pi/40$. The Reynolds number based on the length of ellipse major axis and the incoming velocity is $\text{Re}=200$. The sliding interface is located at $r=1.5$. The computation mesh is plotted in Fig. \ref{ell_mesh} with totally $31516\times2$ elements. 160 nodes are used to discretize the ellipse, and the height of the first layer near the ellipse is $5\times10^{-3}$. The adiabatic non-slip wall is set on the ellipse surface, and the far-field boundary condition is set at the outer boundary.

\begin{figure}[hbt!]
    \centering
    \includegraphics[width=8cm]{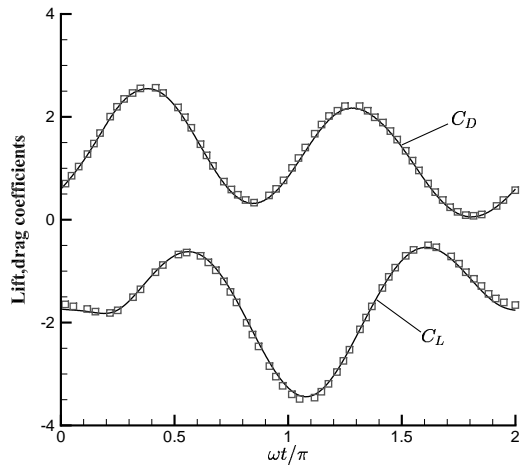}
    \caption{The lift and drag coefficients in one period (Lines: The present numerical result, Symbols: Reference data by Zhang\cite{zhang2015})}\label{clcf}
  \end{figure}

\begin{figure}[hbt!]
      \centering
      \subfigure[$t=0$]{\includegraphics[width=6cm]{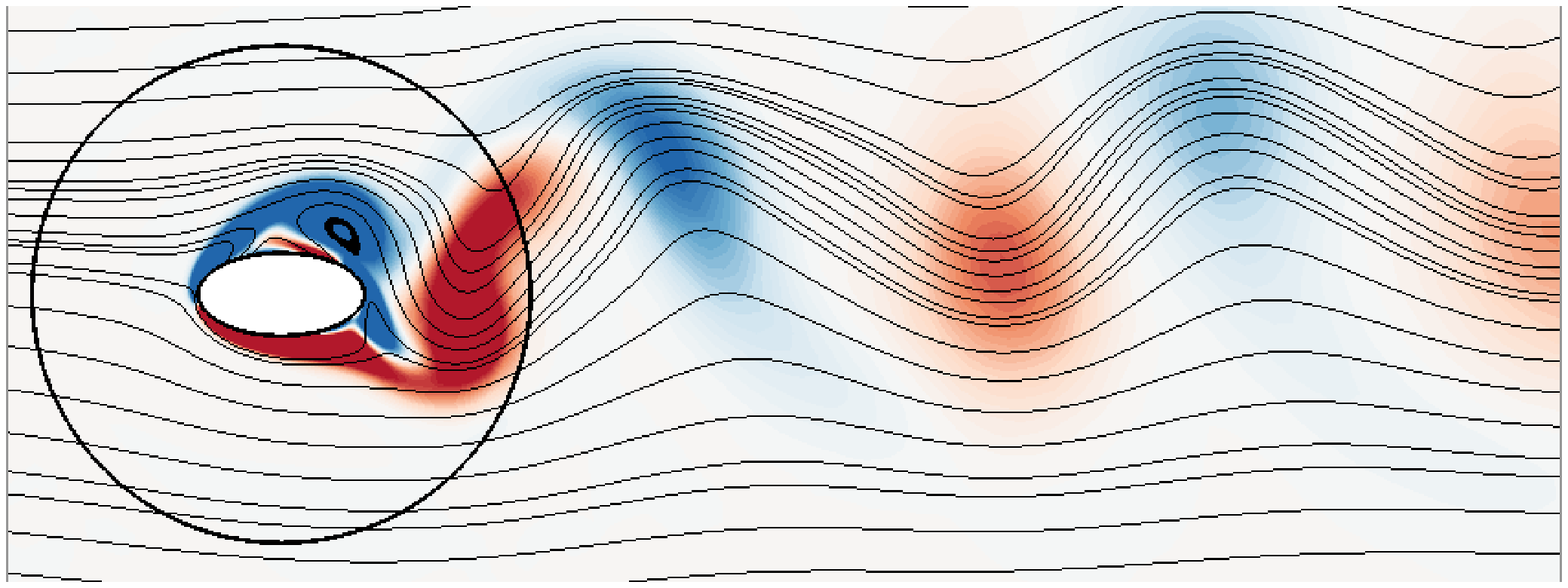}}
      \subfigure[$t=\frac{1}{8}T$]{\includegraphics[width=6cm]{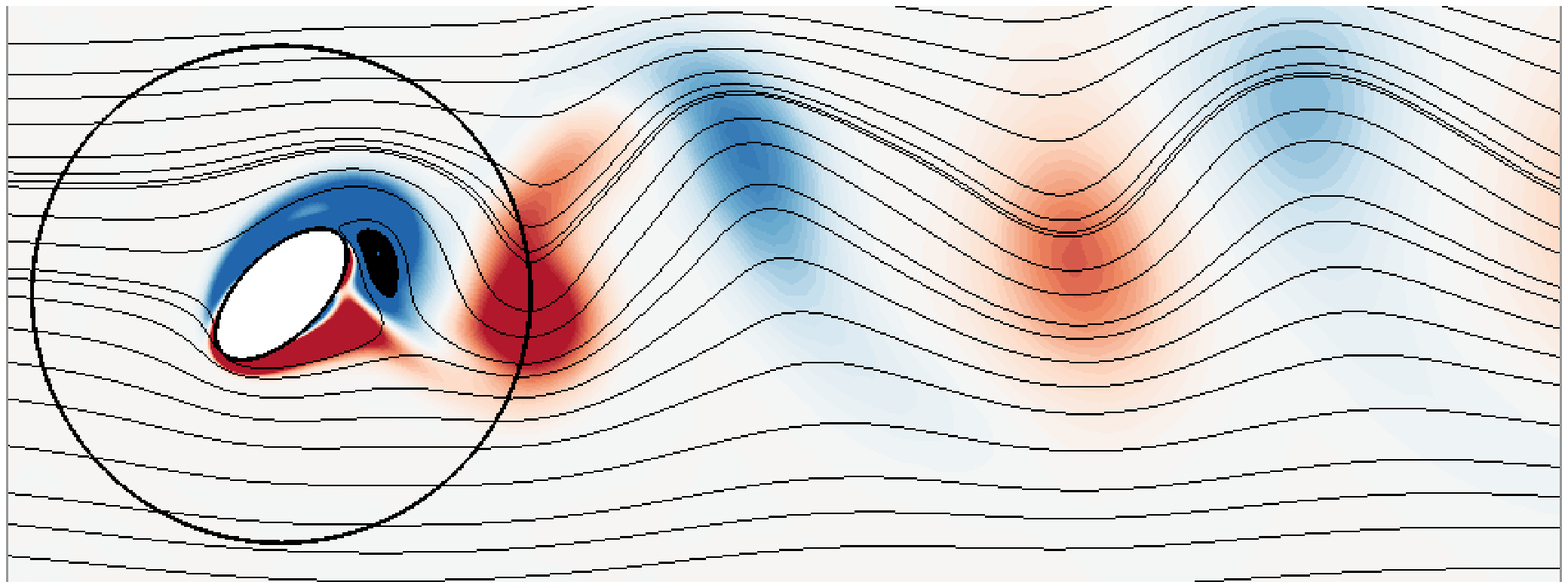}}
      \subfigure[$t=\frac{1}{4}T$]{\includegraphics[width=6cm]{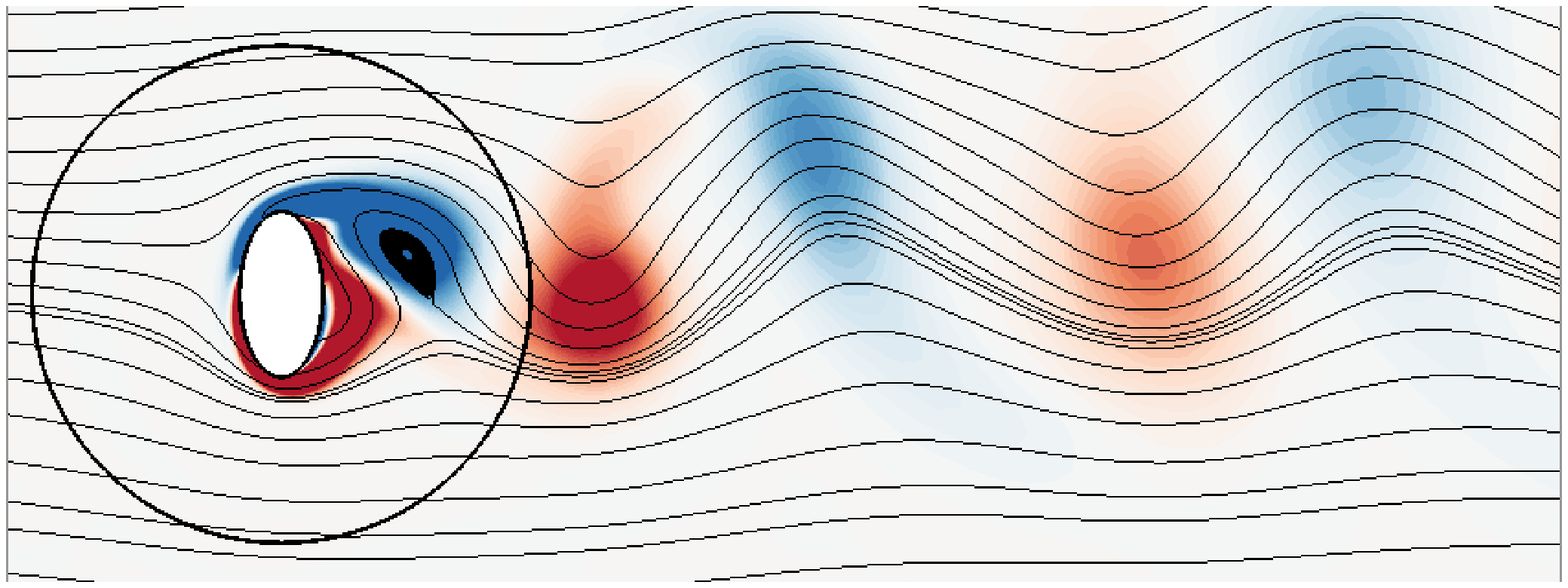}}
      \subfigure[$t=\frac{3}{8}T$]{\includegraphics[width=6cm]{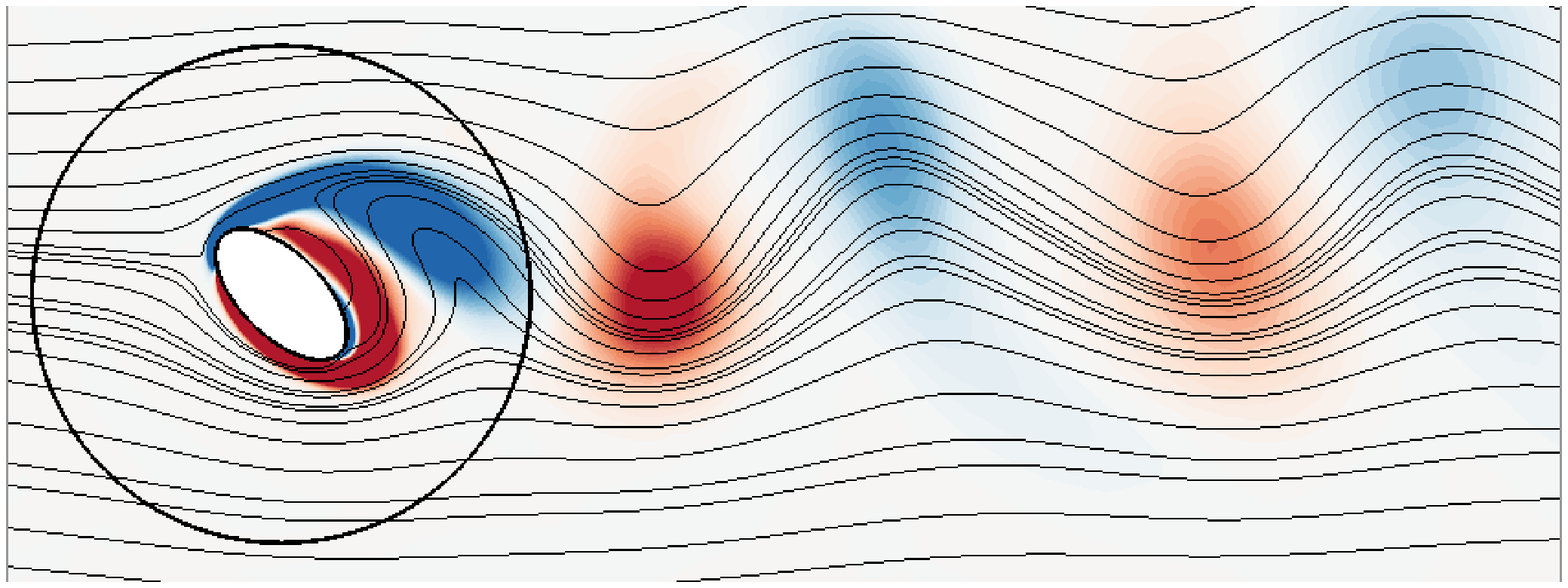}}
      \subfigure[$t=\frac{1}{2}T$]{\includegraphics[width=6cm]{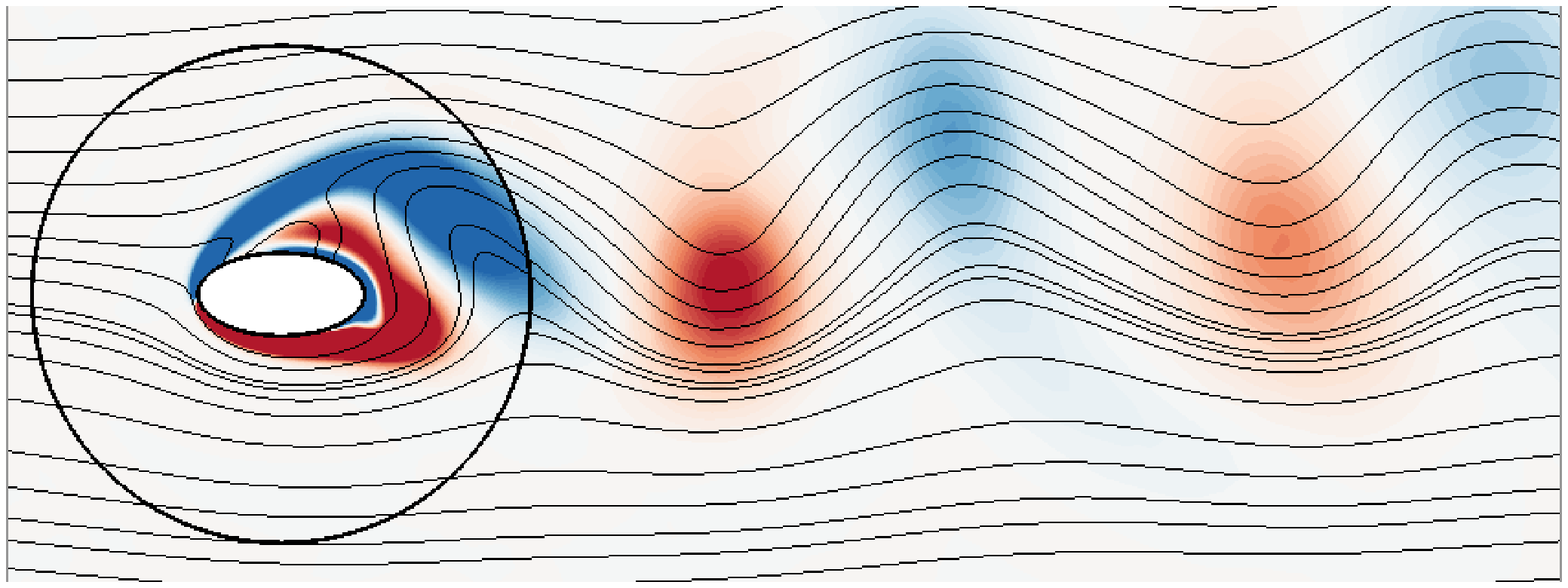}}
      \subfigure[$t=\frac{5}{8}T$]{\includegraphics[width=6cm]{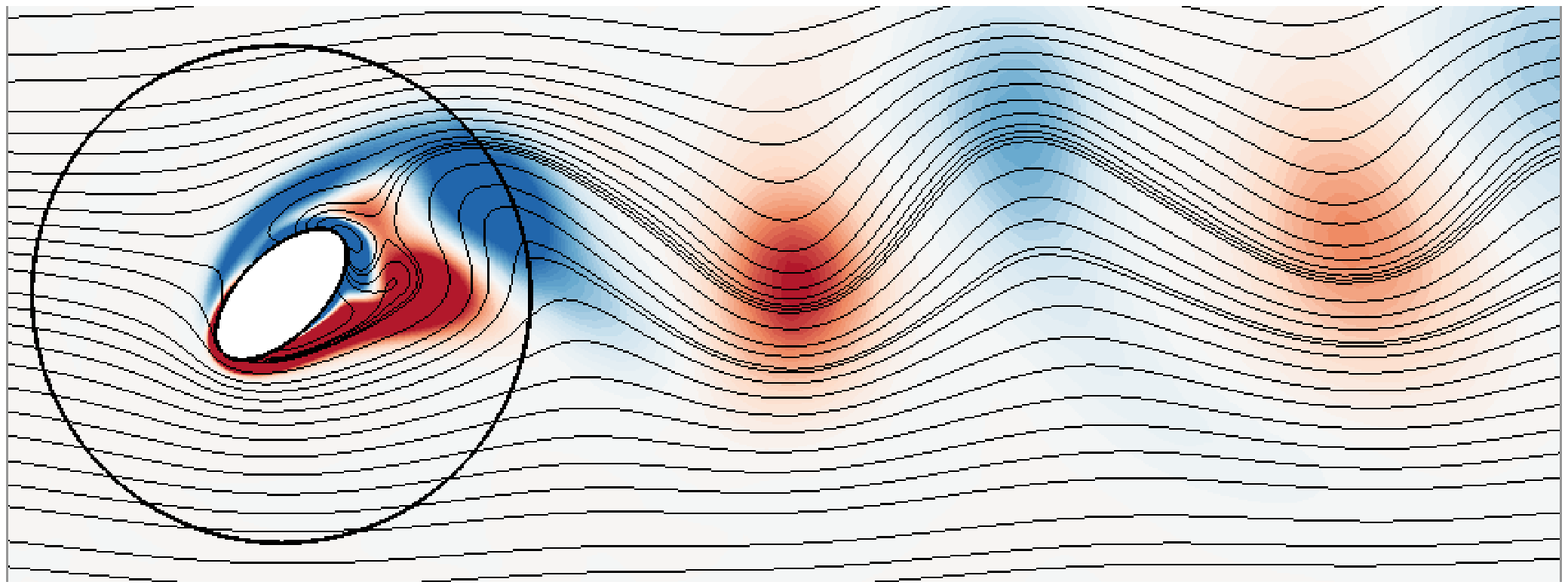}}
      \subfigure[$t=\frac{3}{4}T$]{\includegraphics[width=6cm]{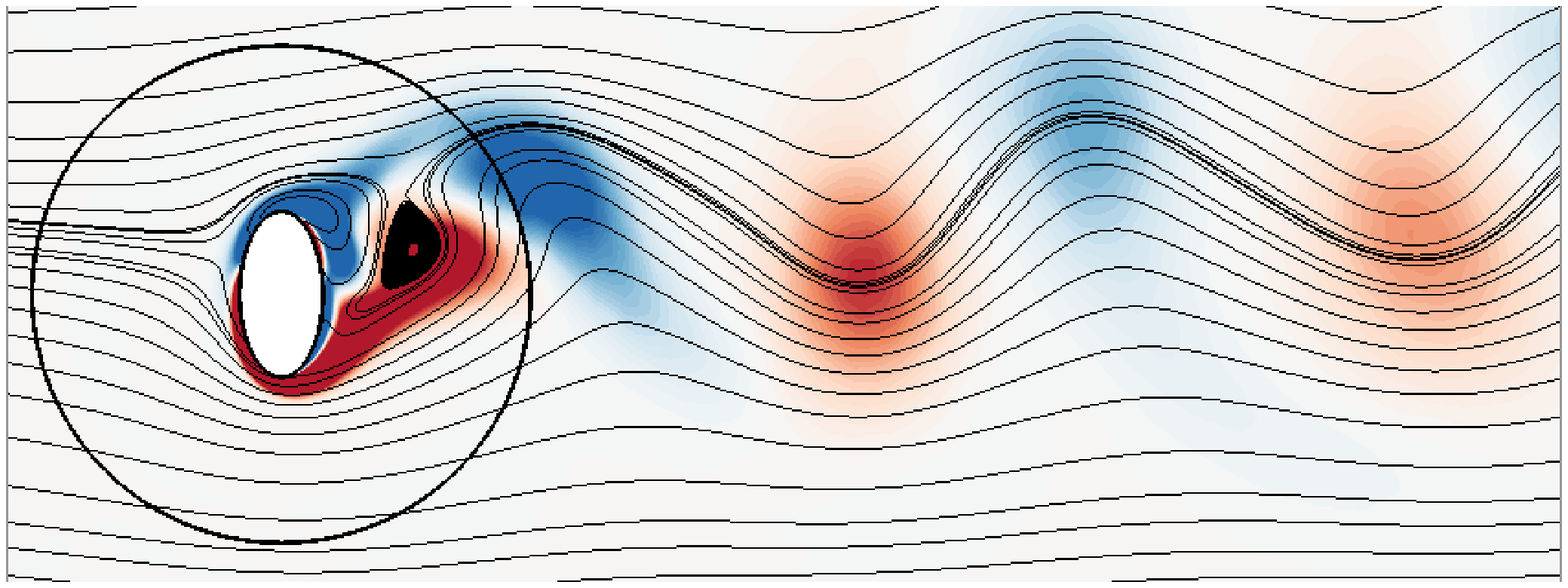}}
      \subfigure[$t=\frac{7}{8}T$]{\includegraphics[width=6cm]{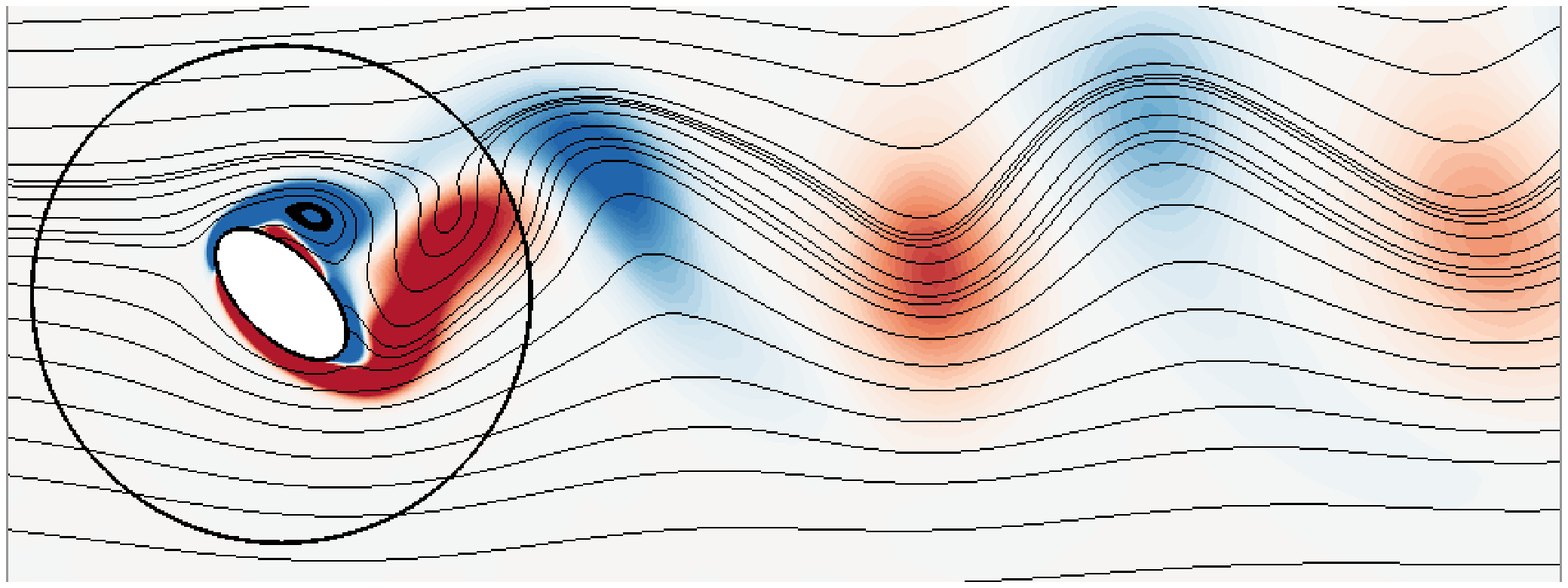}}
        \caption{The vorticity contours and streamlines at different in a periodic}\label{ell_con}
      \end{figure}

  The lift and drag coefficients in one period are shown in Fig. \ref{clcf}.
  The present numerical result is plotted by line, and reference data by Zhang \cite{zhang2015} is shown by symbols. The lift and drag coefficients agree well with the reference data. Also, the vorticity contours and streamlines at different times in one period are plotted in Fig \ref{ell_con}. It can be observed that a clockwise vortex and a counterclockwise vortex generate around the ends of the ellipse. From the time $t=0$ to $t=3/8T$, the clockwise vortex sheds off from the leading edge and hits the trailing edge. While from time $t=1/2T$ to time $t=7/8T$, a counterclockwise vortex slowly emerges and then goes downstream without reattaching to the ellipse. This makes the flow not fully symmetric in a periodic cycle. And the whole process repeats as the ellipse rotates.

\subsection{Stirred tank}
This case is a 2-D laminar case as proposed in \cite{zhang2015}. The computational domain is composed of several parts: a $r=0.5$ cylinder located on the original point, an outer wall with a radius of 5, six uniformly distributed agitating blades with a thickness of 0.1, each extending from $r=1$ to $r=2$, four baffles with same thickness and height 1 installed on the outer wall. The computational domain is split into two parts, an inner rotating part, and an outer fixed part. The sliding interface is located at $r=3$.

\begin{figure}[hbt!]
  \centering
    \includegraphics[width=6cm]{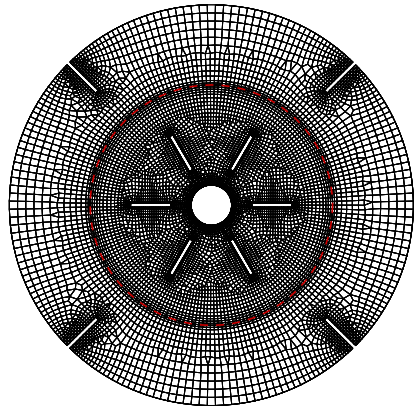}
    \caption{Mesh used in stirred tank case (red dotted lines indicate sliding interfaces)}\label{st_mesh}
  \end{figure}

The initial condition is set as $\rho_0=1.0,p_0=1.0/\gamma, u=v=0$. The inner part with the cylinder and six blades rotates at angular speed $\omega=0.2$, so the Mach number defined by cylinder surface velocity is $M_i=0.05$. And the Reynolds number defined by the diameter of the inner cylinder and the angular speed is $\text{Re}=\rho \omega d^2/\mu=100$. Nonslip wall boundary conditions are applied to all boundaries. The adiabatic wall boundary condition is adopted on the six blades, and the isothermal wall condition is used on other walls. The mesh used in the computation is plotted in Fig. \ref{st_mesh}, with $10242\times 2$ elements.

\begin{figure}[hbt!]
    \centering
    \subfigure[$\omega t=0.25$]{\includegraphics[width=6cm]{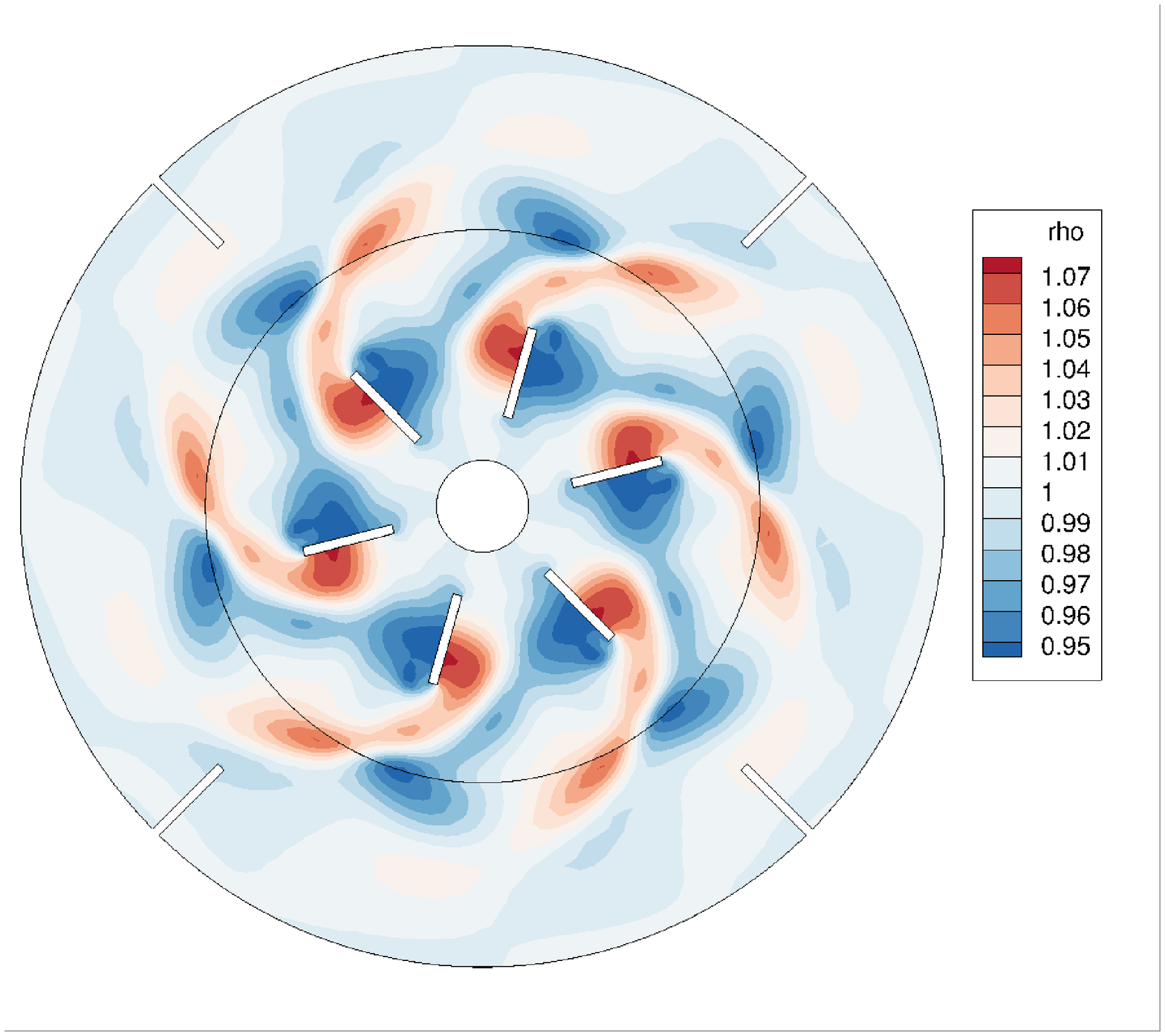}}
    \subfigure[$\omega t=1.5$]{\includegraphics[width=6cm]{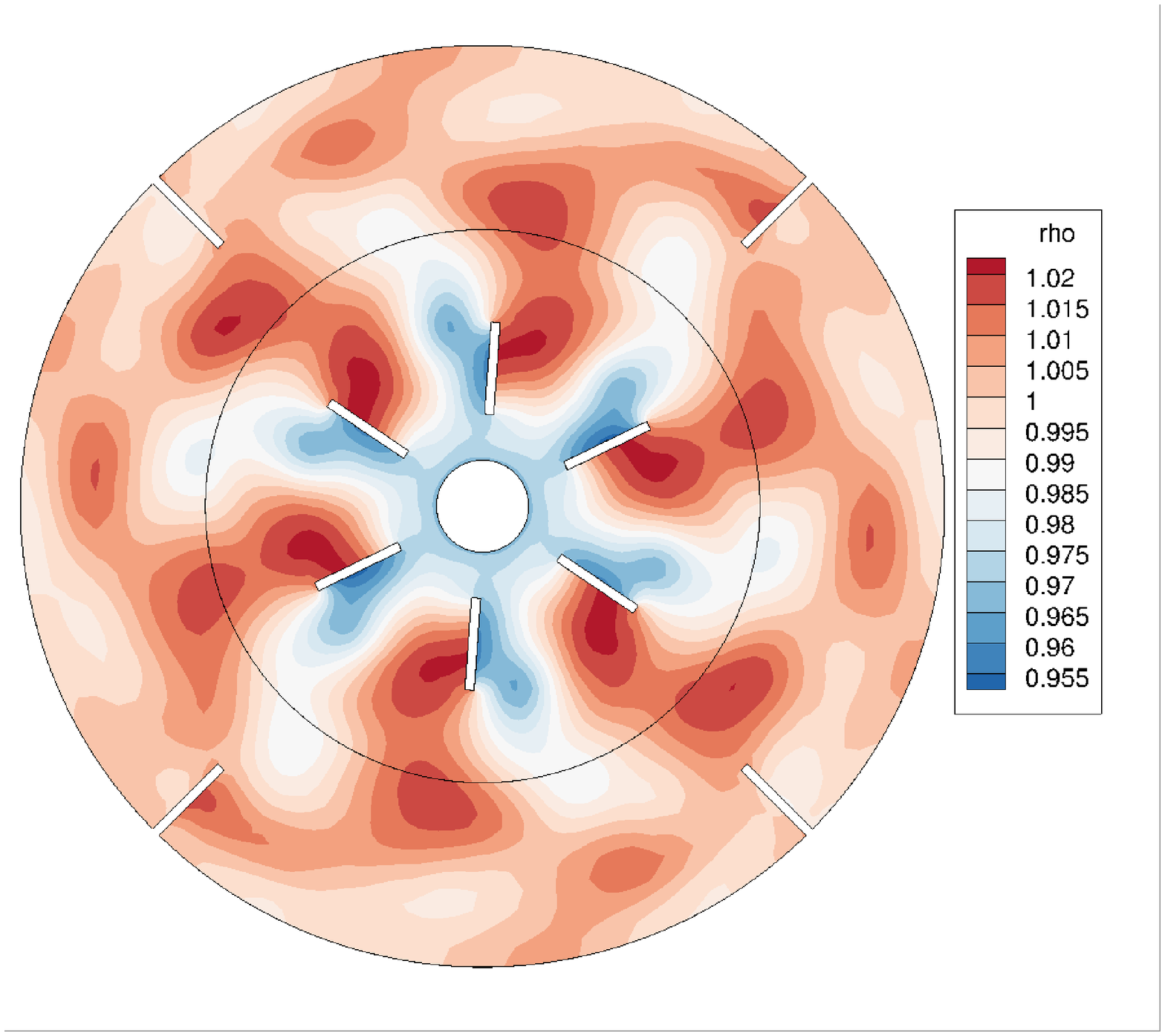}}
    \subfigure[$\omega t=14$]{\includegraphics[width=6cm]{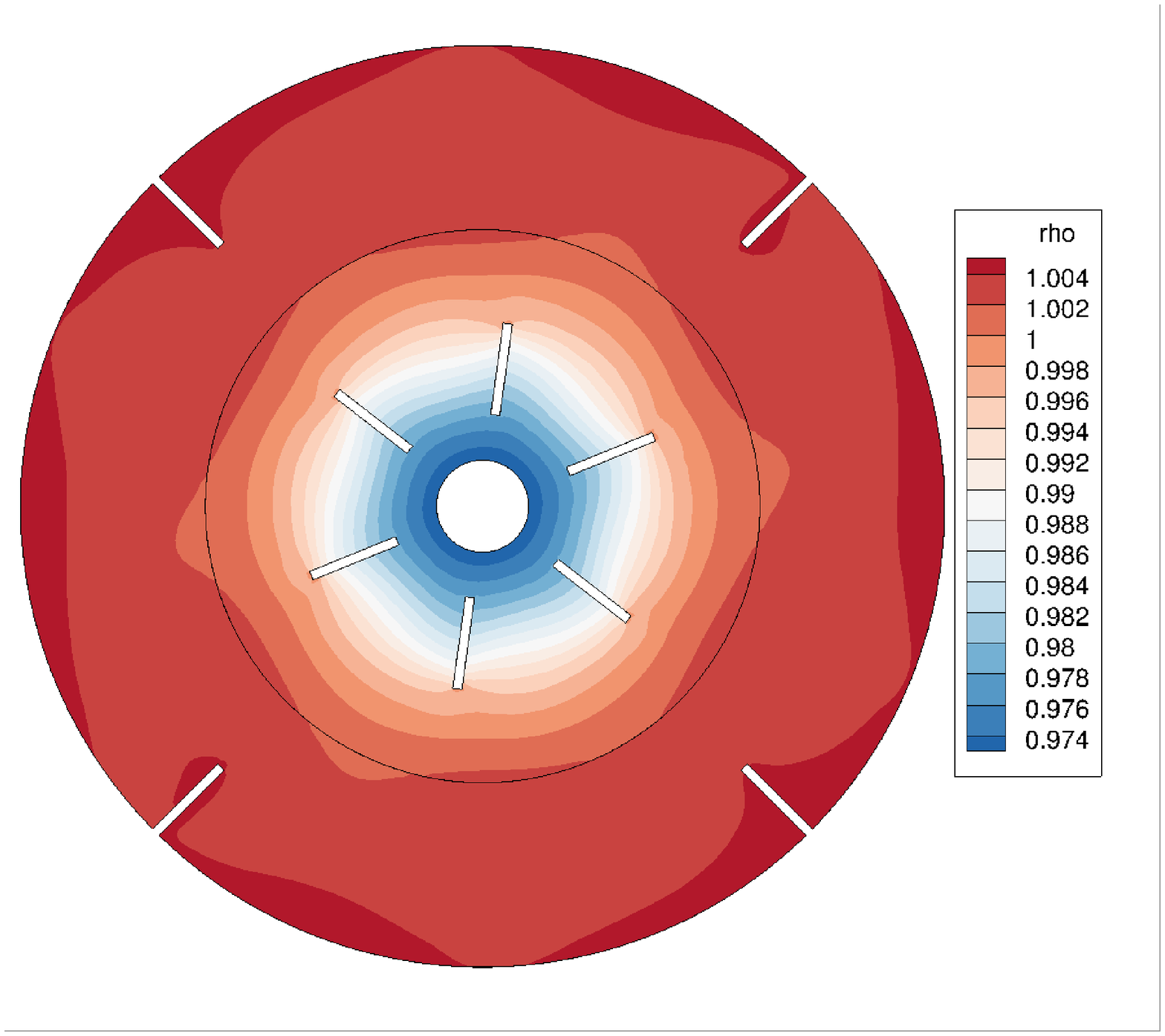}}
    \subfigure[$\omega t=100 $]{\includegraphics[width=6cm]{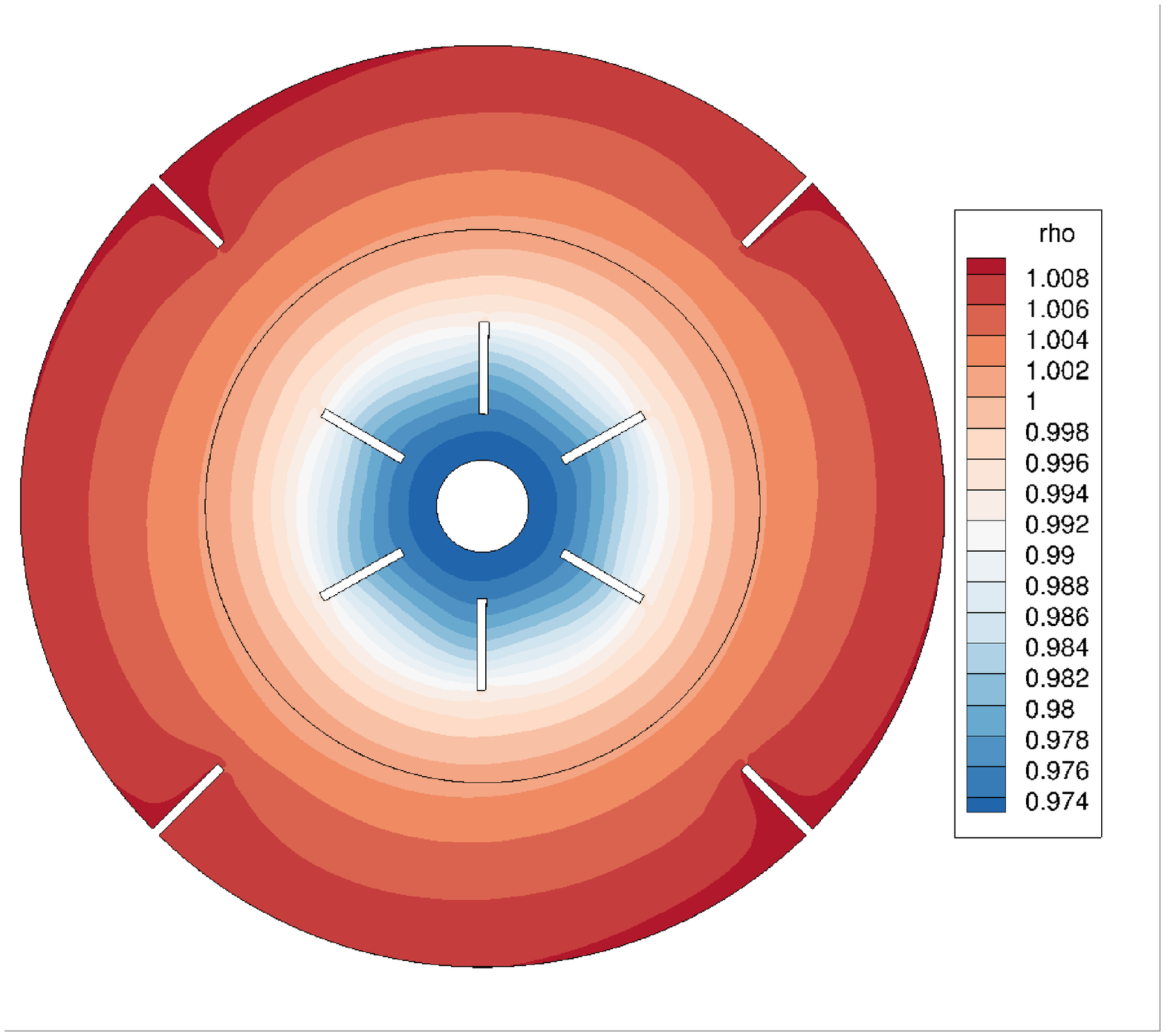}}
      \caption{The density contour of stirred tank}\label{stdens}
    \end{figure}

  The density contours at different times are plotted in Fig. \ref{stdens}. At the time $\omega t =0.25$, the fluid is pushed and squeezed by blades, so large fluctuations can be observed. Soon the fluid becomes very chaotic due to the baffles and the outer wall. At the time $\omega t =1.5$, vortical structures are generated by flow passing the baffles and associated with bouncing pressure waves, blades-induced vortices, unsteady boundary layers, etc. As the blade continues to rotate for a longer time, the chaotic flow structure slowly dissipates, and the flow structure becomes organized. Finally, the flow field reaches a quasi-steady state in the rotating reference framework and changes little with time. The density variation becomes smaller and smaller, and the contour closes to uniform in the circumferential direction. The radial gradient of density is caused by centrifugal force.

\subsection{Ma = 3 cylinder}
\begin{figure}[hbt!]
  \centering
 \includegraphics[width=6cm]{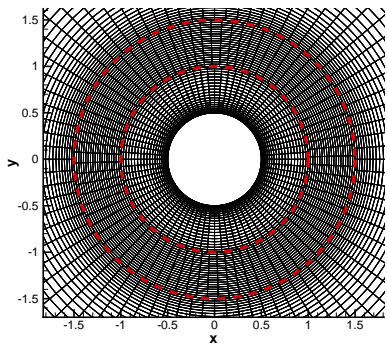}
    \caption{Mesh used in Ma=3 cylinder (red dotted lines indicate sliding interfaces)}\label{cyli_mesh}
  \end{figure}

A steady supersonic flow is used to show the influence of sliding mesh in discontinuous flow. There is a cylinder with radius $r=0.5$ located at center, and the computational domain is a cylindrical domain with radius $R=10$. The whole domain is divided into three parts: from $R_1=0.5$ to $R_2=1$, from $R=1$ to $R=1.5$ and from $R=1.5$ to $R=10$. In the simulation, the second part rotates at angular speed $\omega=1$, so the sliding interfaces are located at $R=1$ and $R=1.5$. And zero angular speed $\omega=0$ case is also calculated for comparison. The inviscid slip wall boundary condition is used on the cylinder surface, and the far-field boundary condition with income Mach number 3 is used on the outer boundary. As shown in Fig. \ref{cyli_mesh}, total $6320\times2$ elements are used in simulation, and red dotted lines indicate the sliding interfaces.

\begin{figure}[hbt!]
    \centering
      \includegraphics[width=6cm]{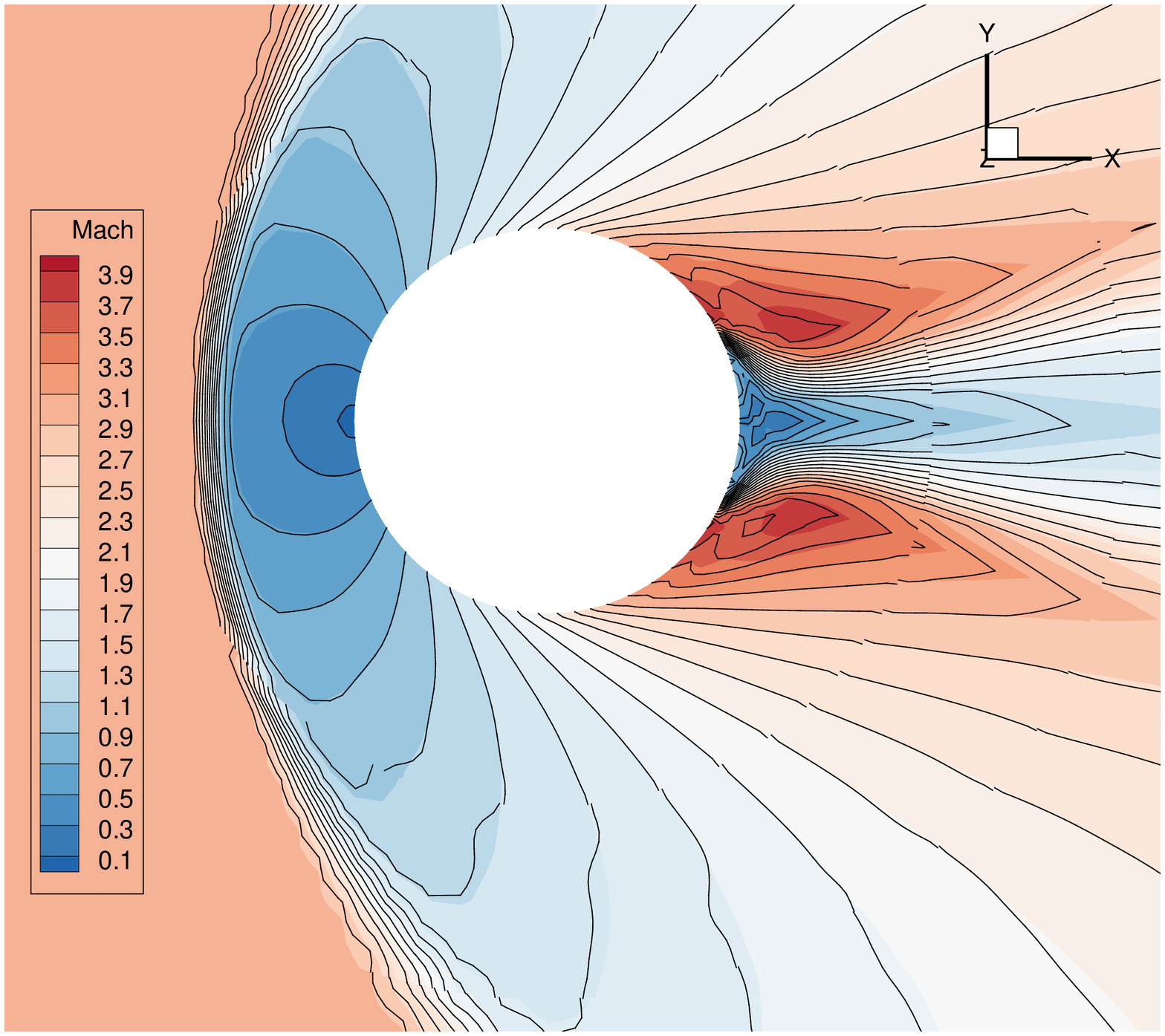}
      \includegraphics[width=6cm]{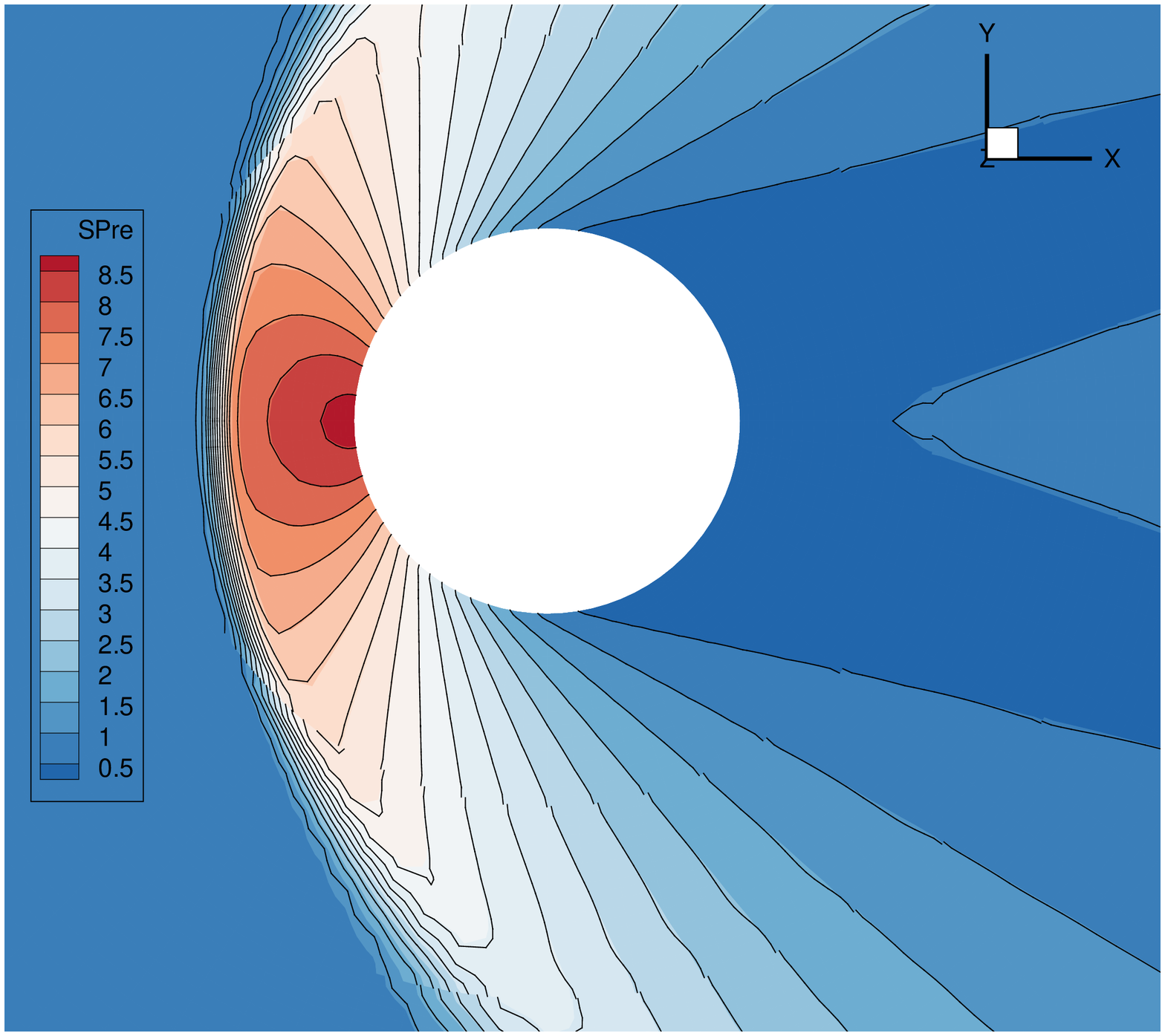}
      \caption{The Mach number and static pressure contours of Ma=3 cylinder (Contours represent the no rotation case and lines represent the rotating case.)}\label{cyli_contour}
    \end{figure}

As shown in Fig. \ref{cyli_contour},
the Mach number and static pressure contours are plotted. No oscillation can be observed near the shock. Overall, no rotating and rotating mesh cases agree well with each other. In the Mach number contours, slight asymmetry can be observed in the wake of the cylinder, where the density and pressure are very low, and the non-linear weights become sensitive to the local geometry.

\subsection{ Three cylinders rotating at supersonic speed}
\begin{figure}[hbt!]
  \centering
  \includegraphics[width=6cm]{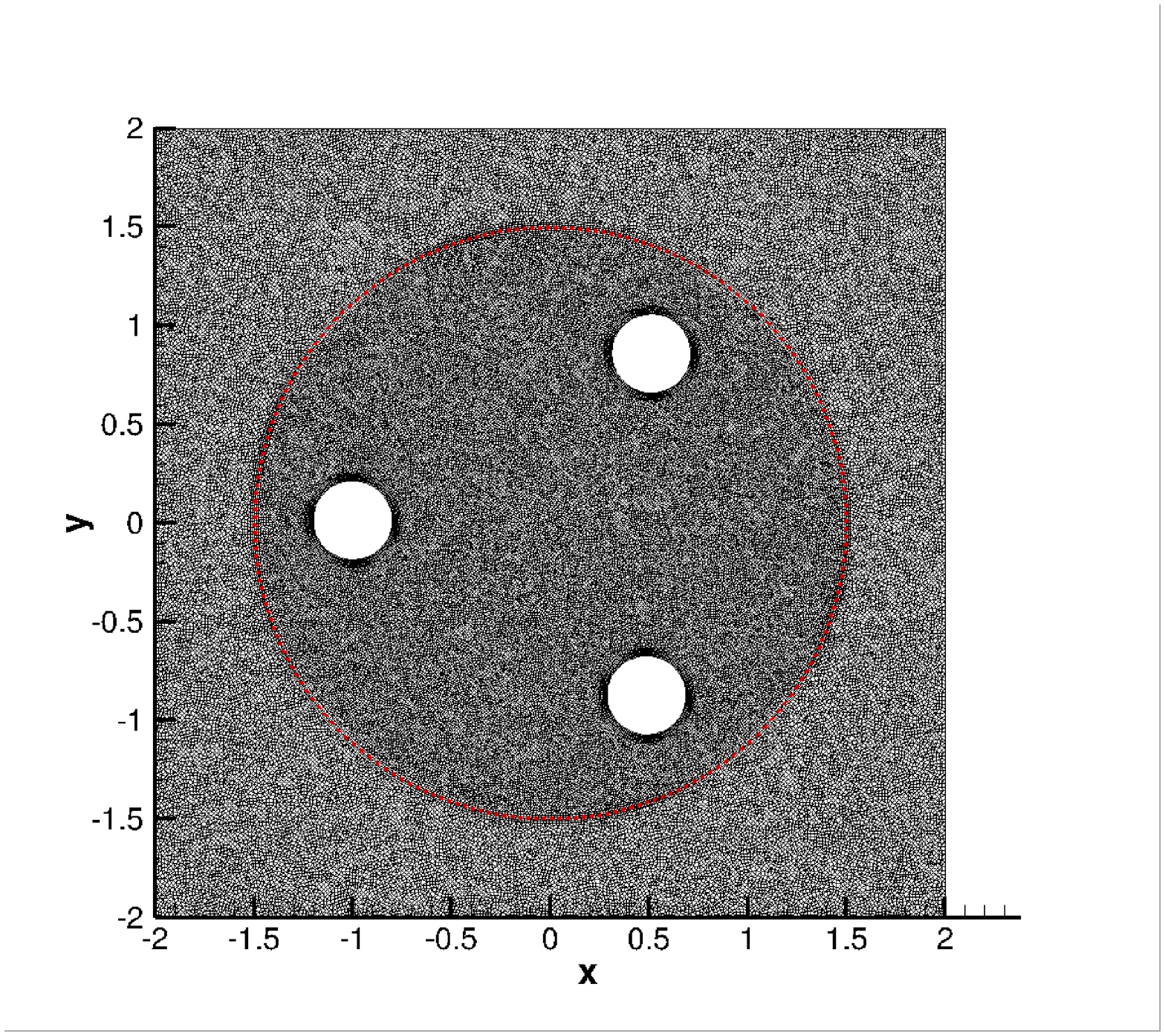}
  \includegraphics[width=6cm]{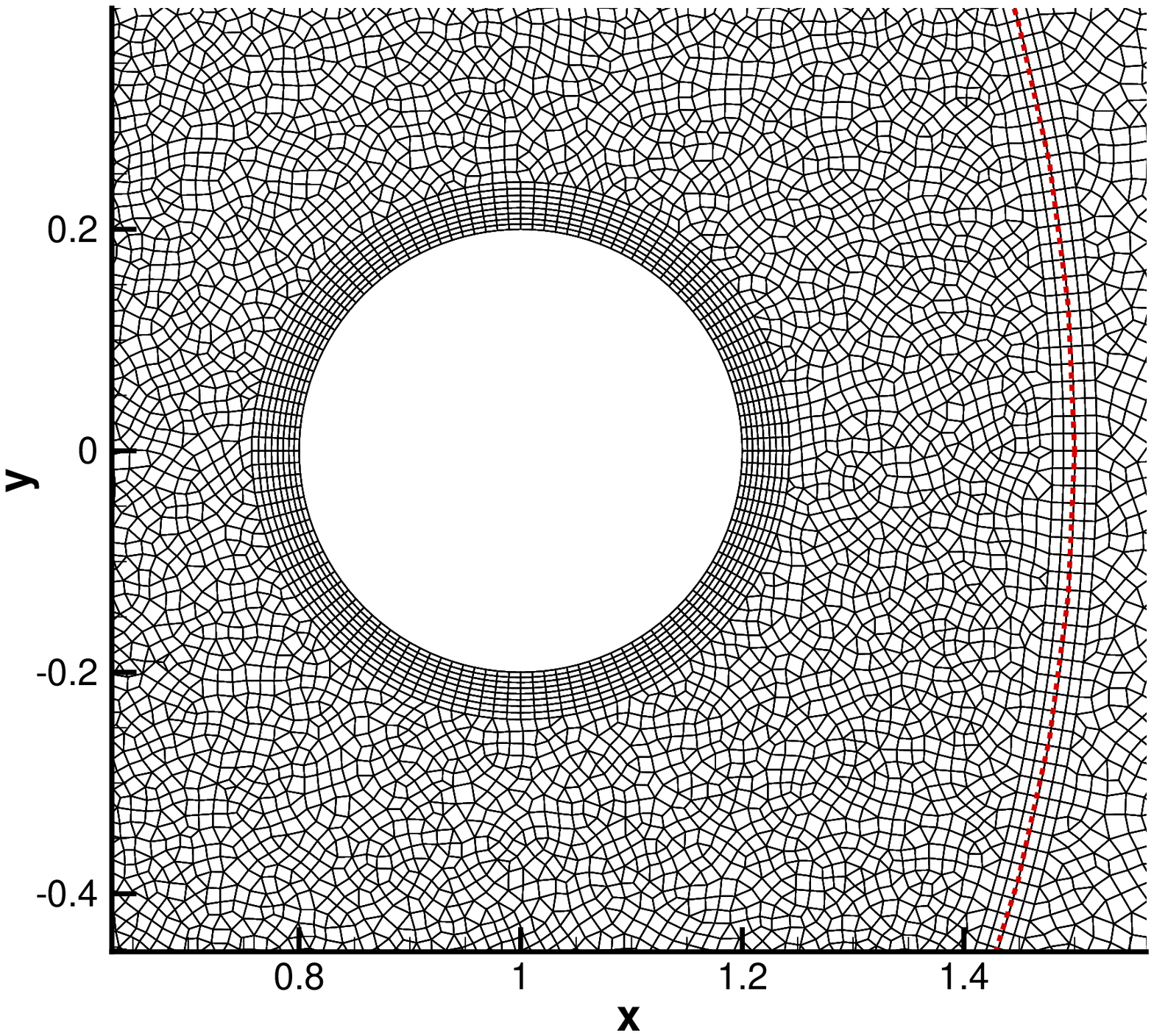}
  \caption{Mesh for three cylinders rotating in a compressible gas at supersonic speed}\label{3cmesh}
\end{figure}

This is an unsteady case with complicated shock interactions, which was used to illustrate the applicability of the diffuse interface model \cite{kemmSimpleDiffuseInterface2020}. The computational domain is $[-2,2]\times[-2,2]$. Three cylinders with radius $r_i=0.2$ are located at position $(x,y)=(R_0\cos{\psi_i},R_0\sin{\psi_i})$, where $R_0=1.0$. These cylinders rotate clockwise at the same angle speed $\omega=3$, which yield a Mach 3 speed at the location of the cylinders. The sliding interface is at the position of $r=1.5$. The periodic boundary condition is applied in the $x$ and $y$ directions. The mesh is plotted in Fig. \ref{3cmesh} and total $90036\times2$ elements are used.

\begin{figure}[hbt!]
   \centering
   \subfigure[$t=0.35$]{\includegraphics[width=6cm]{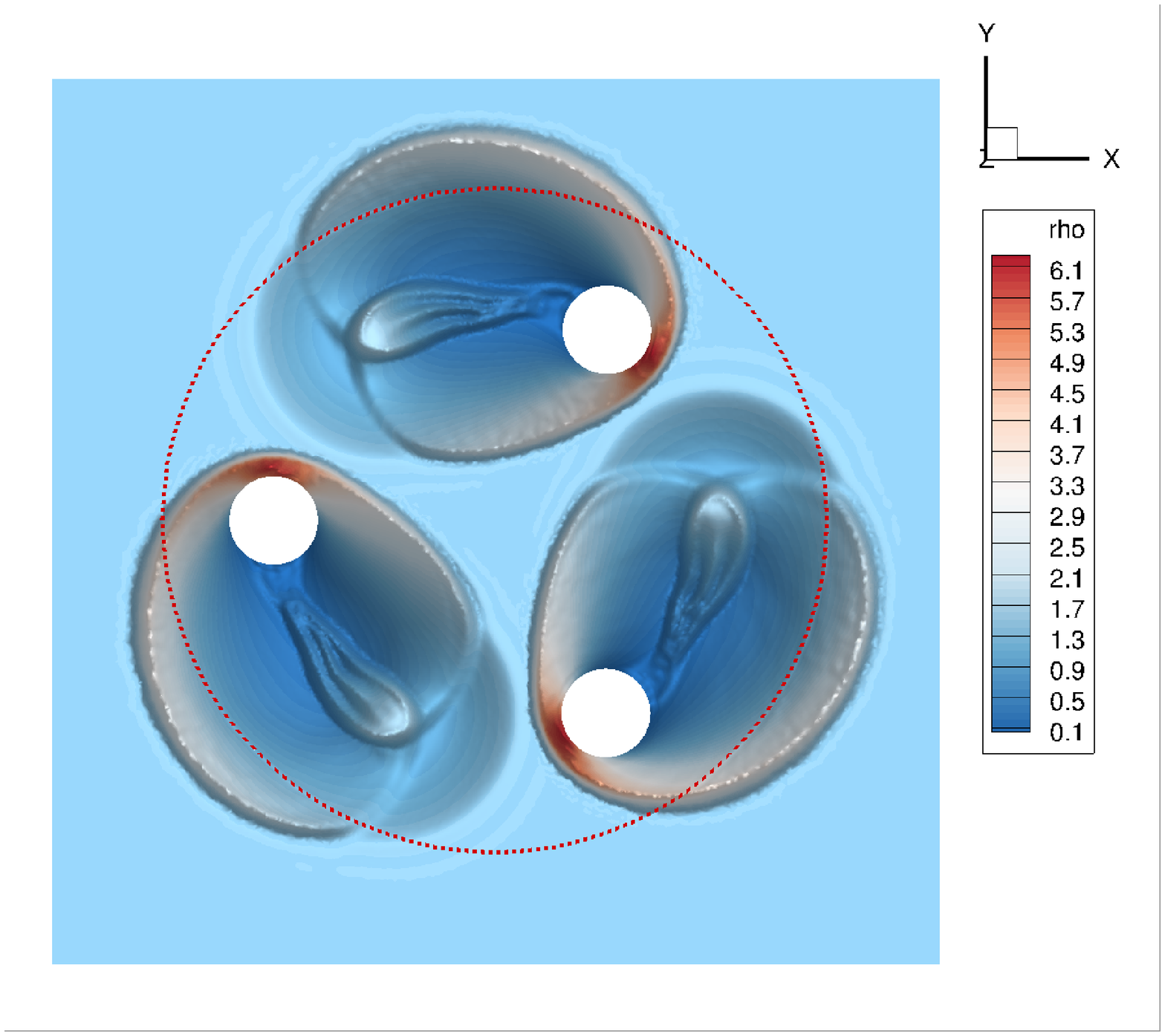}}
   \subfigure[$t=0.5$]{\includegraphics[width=6cm]{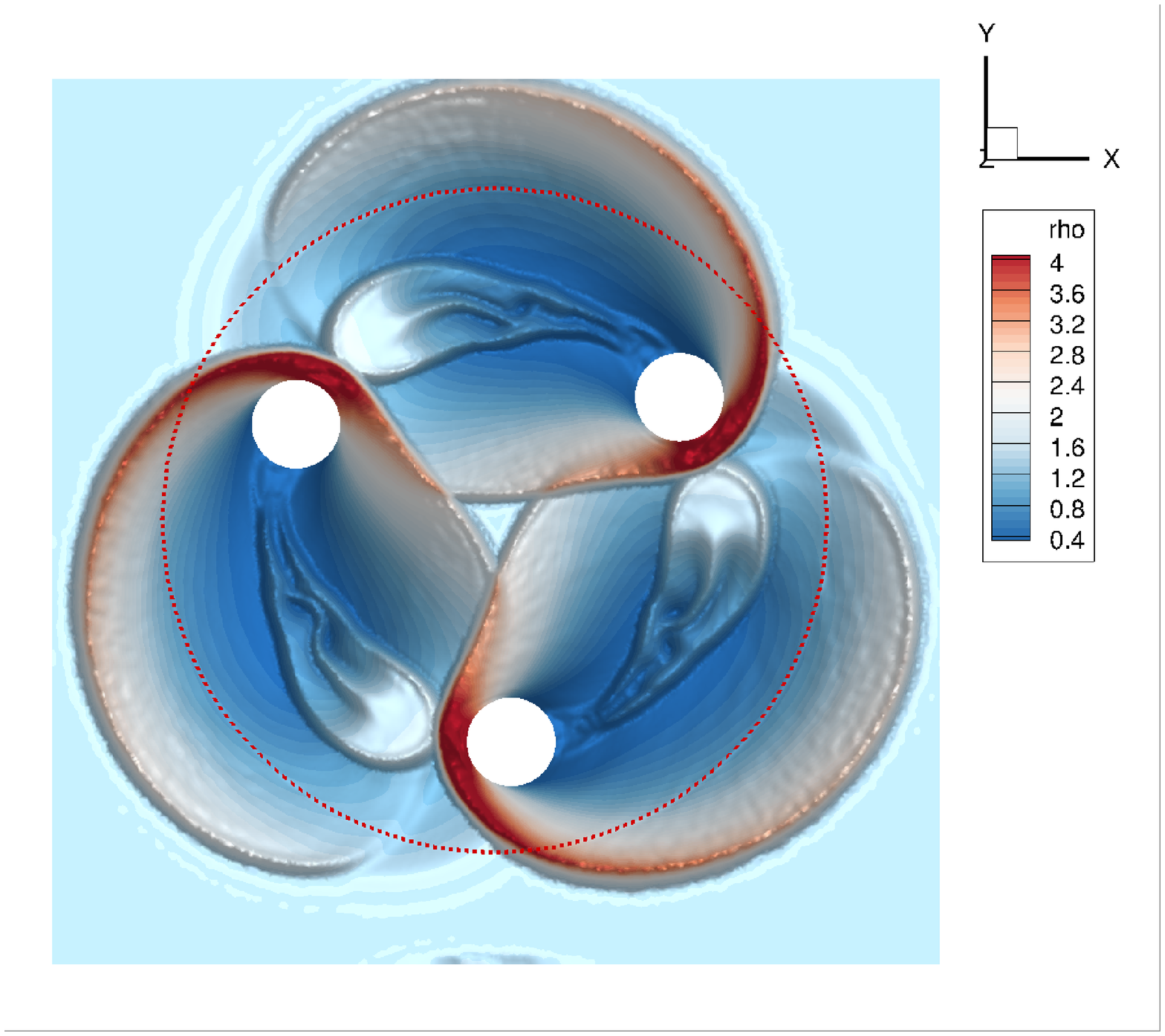}}
   \subfigure[$t=0.75$]{\includegraphics[width=6cm]{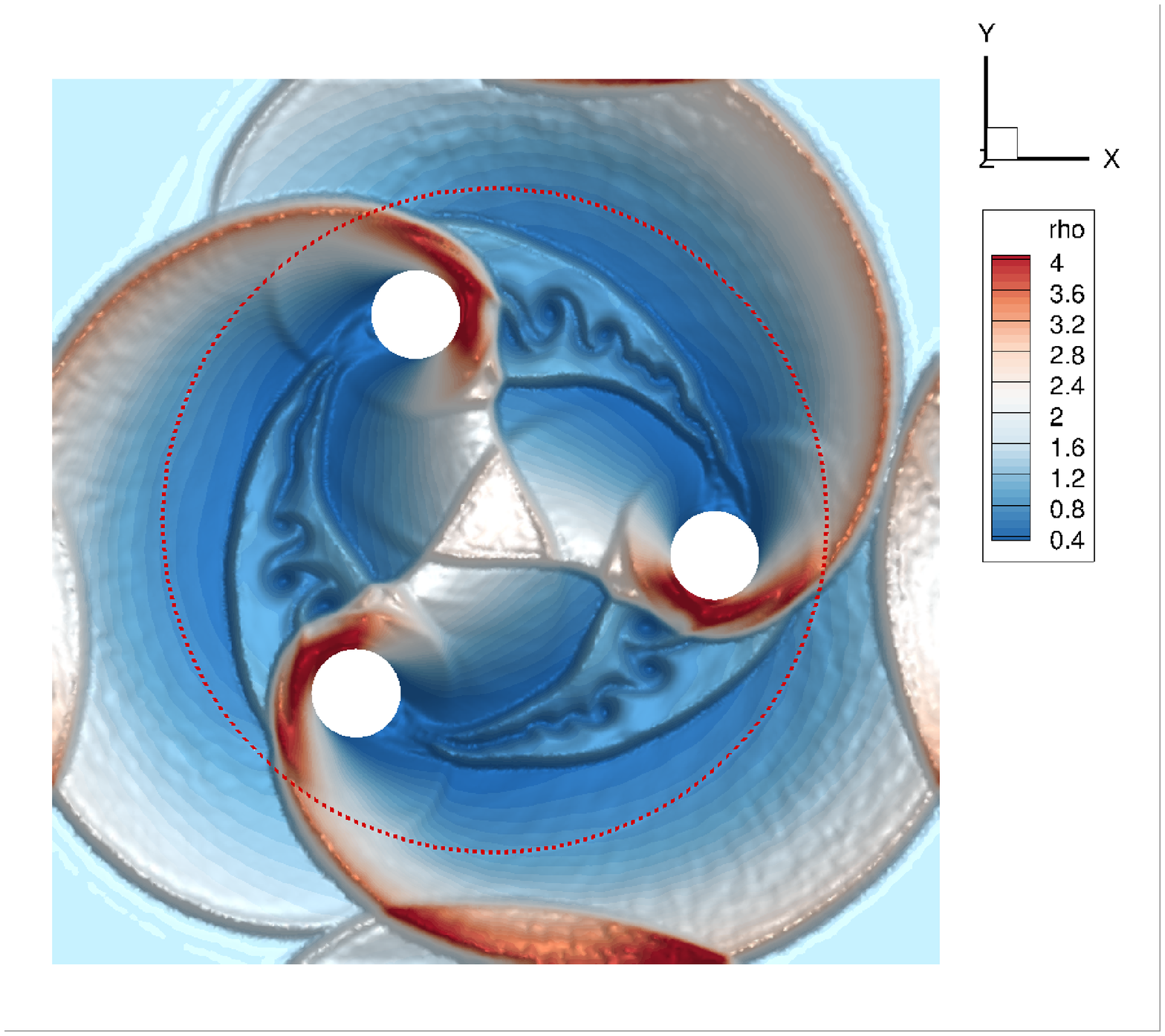}}
   \subfigure[$t=1.75$]{\includegraphics[width=6cm]{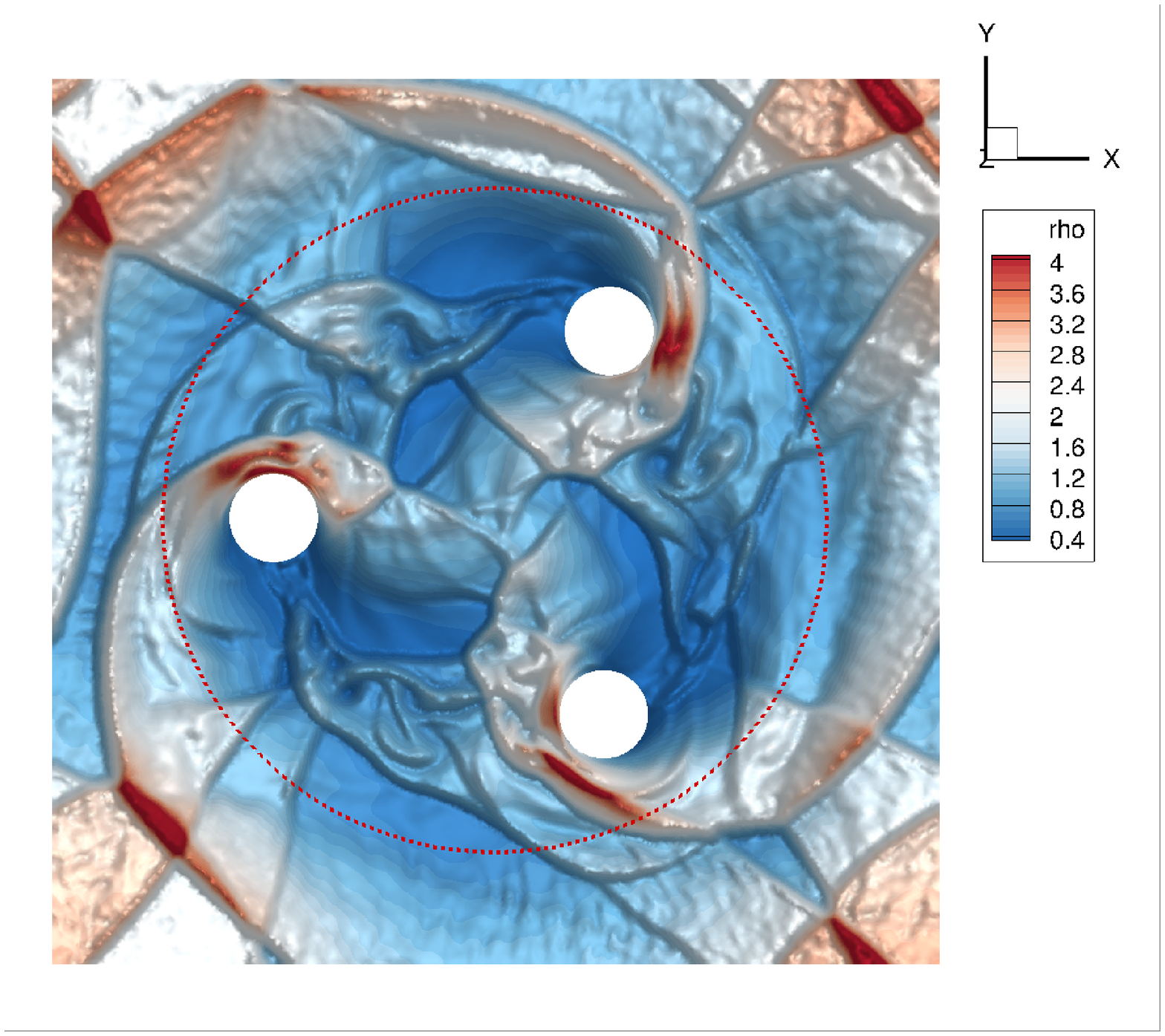}}
     \caption{The density contour of three cylinders rotating in a compressible gas at supersonic rotating speed (the red dotted cycle indicates sliding interface)}\label{3cdens}
   \end{figure}

The density contours are shown in Fig. \ref{3cdens}. The shocks emerge in the front of cylinders and then interact with the trailing wake. No unphysical oscillation can be observed on the sliding interface, which indicates that the proposed method can deal with moving shocks well.

   \section{Conclusion}
   In this paper, a third-order CGKS is developed in rotating coordinate frame with the combination of sliding mesh method for simulating flow
   problems with rotating parts.
   Due to the kinetic nature of the gas evolution model, the dynamic effect from centrifugal and Coriolis forces in the rotating frame can be easily incorporated into the time accurate flux function and flow variable updates at a cell interface.
   At a result, both cell-averaged flow variables and their gradients can be updated and used in the high-order compact reconstruction.
   The high-order and compactness of the scheme have advantages for flow simulation with rotating parts in capturing the unsteady flow passing through the sliding interface.
   The current CGKS can use a large CFL number, such as CFL number 0.5, in the determination of time step in the flow simulation with highly compressible shock wave.
   Many test cases, covering viscid and inviscid, subsonic and supersonic cases, are used to validate the scheme.
   The numerical performance of the scheme in the density wave propagation, vortex flow, shock passing through sliding interfaces, and rotating cylinders at supersonic speed, shows the accuracy and robustness of the high-order method.
   The current scheme can be extended straightforwardly to the three-dimensional case.
   In the coming work, large-scale three dimensional flow computations, such as propeller noise and wake-shock interactions in the transonic compressor, will be presented. At the same time, the parallel technology will be further developed to improve computational efficiency in 3D applications. The high efficiency of the scheme can be easily realized because of the compactness of the stencils.

  \appendix
  \section{ Chapman-Enskog Expansion of BGK equation in rotating framework}
  \label{C-E}
  The BGK equation in Eq. (\ref{BGK}) can be written in this form
  \begin{equation*}
  f = g-\tau\left(\frac{\partial f}{\partial t}+w_l\frac{\partial f}{\partial x_l}-\epsilon_{lij}\Omega_iv_j\frac{\partial f}{\partial v_l}\right).
 \end{equation*}
 The formal solution of $f$ can be expanded as
\begin{equation*}
  f=f_0+\epsilon f_1 + \epsilon^2 f_2+ \epsilon^3 f_3+\cdots.
\end{equation*}
By setting $\tau=\epsilon \hat{\tau} $ into the BGK equation directly, we have
  \begin{equation*}
  f = g-\epsilon \hat{\tau}\left(\frac{\partial f}{\partial t}+w_l\frac{\partial f}{\partial x_l}-\epsilon_{lij}\Omega_iv_j\frac{\partial f}{\partial v_l}\right).
 \end{equation*}
An expression of this equation in powers of $\epsilon$ is
   \begin{equation}\label{ce2}
    \begin{aligned}
      f = g&-\epsilon \hat{\tau}\left(\frac{\partial }{\partial t} +w_l\frac{\partial}{\partial x_l}-\epsilon_{lij}\Omega_iv_j\frac{\partial }{\partial v_l}\right)g \\
      &+ \epsilon^2\hat{\tau}\left(\frac{\partial }{\partial t}+w_l\frac{\partial}{\partial x_l}-\epsilon_{lij}\Omega_iv_j\frac{\partial }{\partial v_l}\right) \left[  \hat{\tau}\left(\frac{\partial }{\partial t}+w_k\frac{\partial}{\partial x_k}-\epsilon_{kmn}\Omega_mv_n\frac{\partial }{\partial v_k}\right)g\right]+o(\epsilon^3).
    \end{aligned}
 \end{equation}
  With the implementation of the compatibility condition, after dividing both sides of the equation  by $\epsilon\hat{\tau}$ the moments of Eq. (\ref{ce2}) become
 \begin{equation}\label{cemen}
  \begin{aligned}
    &\int \psi_\alpha \left(\frac{\partial }{\partial t}+w_l\frac{\partial}{\partial x_l}-\epsilon_{lij}\Omega_iv_j\frac{\partial }{\partial v_l}\right)g  \td\boldsymbol{v}\td\Xi \\
    =&\epsilon\int \psi_\alpha\left(\frac{\partial }{\partial t}+w_l\frac{\partial}{\partial x_l}-\epsilon_{lij}\Omega_iv_j\frac{\partial }{\partial v_l}\right) \left[  \hat{\tau}\left(\frac{\partial }{\partial t}+w_k\frac{\partial}{\partial x_k}-\epsilon_{kmn}\Omega_mv_n\frac{\partial }{\partial v_k}\right)g\right] \td\boldsymbol{v}\td\Xi+o(\epsilon^2).
  \end{aligned}
\end{equation}

Defining $L_\alpha=\int \psi_\alpha \left(\frac{\partial }{\partial t}+w_l\frac{\partial}{\partial x_l}-\epsilon_{lij}\Omega_iv_j\frac{\partial }{\partial v_l}\right)g  \td\boldsymbol{v}\td\Xi$ and considering
\begin{equation*}
\begin{aligned}[c]
  &\epsilon_{lij}\Omega_i\int v_j\psi_\alpha \frac{\partial g}{\partial v_l} \td\boldsymbol{v}\td\Xi
  \\ =&\epsilon_{lij}\Omega_i\int \frac{\partial gv_j\psi_\alpha}{\partial v_l} -g\frac{\partial v_j\psi_\alpha}{\partial v_l} \td\boldsymbol{v}\td\Xi
   =\epsilon_{lij}\Omega_i\int -g( \delta_{jl}\psi_\alpha +v_j\frac{\partial \psi_\alpha}{\partial v_l}) \td\boldsymbol{v}\td\Xi \\
   =&-\epsilon_{lij}\Omega_i\int g v_j\frac{\partial \psi_\alpha}{\partial v_l} \td\boldsymbol{v}\td\Xi =-\epsilon_{lij}\Omega_i<v_j\frac{\partial \psi_\alpha}{\partial v_l}>,
\end{aligned}
\end{equation*}
the $L_\alpha$ becomes
\begin{equation*}
     L_\alpha =<\psi_\alpha>_{,t} +<\psi_\alpha w_l>_{,l}+\epsilon_{lij}\Omega_i<v_j\frac{\partial \psi_\alpha}{\partial v_l}>=o(\epsilon).
  \end{equation*}
We can get
\begin{equation*}
  L_1=\rho_t+(\rho W_l)_{,l},
\end{equation*}
for $a=2,3,4$
\begin{equation*}
  L_a=(\rho V_a)_{,t}+(\rho V_aW_l)_{,l}+p_{,a}+\rho\epsilon_{alj}\Omega_lV_j ,
\end{equation*}
and
\begin{equation*}
  L_5=\left(\frac{1}{2}\rho V_n^2+\frac{K+3}{2}p \right)_{,t}+\left(\frac{1}{2}\rho W_k V_n^2+\frac{K+5}{2}pW_k \right)_{,k} +U_kp_{,k}.
\end{equation*}
Then, we have
\begin{equation}\label{oep}
  \begin{aligned}[c]
    \rho V_{a,t}+\rho W_lV_{a,l}+p_{,a}+\rho\epsilon_{alj}\Omega_lV_j =o(\epsilon),\\
    \frac{K+3}{2}\left(p_t +W_kp_{,k} \right)+\frac{K+5}{2}pW_{k,k}=o(\epsilon).
  \end{aligned}
  \end{equation}
With $\hat{\tau}=\hat{\tau}(t,x_i)$, and $R_\alpha$ and $S_\alpha$ defined as
\begin{equation*}
\begin{aligned}[c]
  R_\alpha = &\int \psi_\alpha\left(\frac{\partial }{\partial t}+w_l\frac{\partial}{\partial x_l}-\epsilon_{lij}\Omega_iv_j\frac{\partial }{\partial v_l}\right) \left[  \hat{\tau}\left(\frac{\partial }{\partial t}+w_k\frac{\partial}{\partial x_k}-\epsilon_{kmn}\Omega_mv_n\frac{\partial }{\partial v_k}\right)g\right] \td\boldsymbol{v}\td\Xi,
\end{aligned}
\end{equation*}
\begin{equation*}
  \begin{aligned}[c]
    S_\alpha = &-\int \psi_\alpha\epsilon_{lij}\Omega_iv_j\frac{\partial }{\partial v_l} \left[  \hat{\tau}\left(\frac{\partial }{\partial t}+w_k\frac{\partial}{\partial x_k}-\epsilon_{kmn}\Omega_mv_n\frac{\partial }{\partial v_k}\right)g\right] \td\boldsymbol{v}\td\Xi\\
    = &\epsilon_{lij}\Omega_i\int( \psi_\alpha\frac{\partial v_j}{\partial v_l} +v_j\frac{\partial\psi_\alpha }{\partial v_l}) \left[  \hat{\tau}\left(\frac{\partial }{\partial t}+w_k\frac{\partial}{\partial x_k}-\epsilon_{kmn}\Omega_mv_n\frac{\partial }{\partial v_k}\right)g\right] \td\boldsymbol{v}\td\Xi \\
    = &\epsilon_{lij}\Omega_i\int v_j\frac{\partial\psi_\alpha }{\partial v_l} \left[  \hat{\tau}\left(\frac{\partial }{\partial t}+w_k\frac{\partial}{\partial x_k}-\epsilon_{kmn}\Omega_mv_n\frac{\partial }{\partial v_k}\right)g\right] \td\boldsymbol{v}\td\Xi,
  \end{aligned}
  \end{equation*}
we have $S_1=S_5=0$ for $\alpha =1$ or $\alpha=5$, and $S_\alpha=-\hat{\tau}\epsilon_{\alpha ij}\Omega_iL_j = o(\epsilon)$ for $\alpha=2,3,4$.
So, the $R_\alpha$ becomes
\begin{equation*}
  \begin{aligned}[c]
    R_\alpha = &(\hat{\tau}L_\alpha)_{,t} +S_\alpha + \frac{\partial}{\partial x_l}\int \hat{\tau}w_l\psi_\alpha   \left(\frac{\partial }{\partial t}+w_k\frac{\partial}{\partial x_k}-\epsilon_{kmn}\Omega_mv_n\frac{\partial }{\partial v_k}\right)g\td\boldsymbol{v}\td\Xi \\
    = & \frac{\partial}{\partial x_l}\int \hat{\tau}v_l\psi_\alpha   \left(\frac{\partial }{\partial t}+w_k\frac{\partial}{\partial x_k}-\epsilon_{kmn}\Omega_mv_n\frac{\partial }{\partial v_k}\right)g\td\boldsymbol{v}\td\Xi -(u_l\hat{\tau}L_\alpha)_{,l} +o(\epsilon) \\
    =&\left\{\hat{\tau}\left[ <\psi_\alpha v_l>_{,t} +<\psi_\alpha  v_lw_k >_{,k}   +\epsilon_{kmn}\Omega_m(<v_n\psi_\alpha>\delta_{lk}+<v_nv_l\frac{\partial \psi_\alpha}{\partial v_k}> )\right] \right\}_{,l}+o(\epsilon)
  \end{aligned}
  \end{equation*}
Firstly, for $\alpha=1$
\begin{equation*}
  R_1= (\hat{\tau}L_l  )_{,l}+o(\epsilon)=o(\epsilon).
\end{equation*}
The equation becomes
\begin{equation*}
  \rho_{,t}+(\rho W_l)_{,l}=o(\epsilon^2).
\end{equation*}
For $a=2,3,4$
\begin{equation*}
\begin{aligned}[c]
  R_a
 =
 & \left\{\hat{\tau}\left[ <v_av_l>_{,t} +<v_a v_lw_k >_{,k}   +\epsilon_{kmn}\Omega_m(<v_nv_a>\delta_{lk}+<v_nv_l>\delta_{ak} )\right] \right\}_{,l}+o(\epsilon),
\end{aligned}
\end{equation*}
with the consideration of equation (\ref{oep}), we can get
\begin{equation*}
    R_a =\left\{\hat{\tau}\left[p  (V_{a, l}+V_{l,a}-\frac{2}{3}V_{k,k}\delta_{al})+\frac{2}{3}\frac{K}{K+3}pV_{k,k}\delta_{al}    \right] \right\}_{,l}+o(\epsilon).
  \end{equation*}
And the method can be used for $\alpha=5$
  \begin{equation*}
    \begin{aligned}[c]
      R_5 =
       &\left\{\hat{\tau}\left[<\psi_5 v_l>_{,t}+<\psi_5 v_lw_k >_{,k}  +\epsilon_{lmn}\Omega_m<v_n\psi_5>\right] \right\}_{,l}+o(\epsilon)\\
       =& \left\{\hat{\tau}\left[ \frac{K+5}{2}p\left(\frac{p}{\rho}\right)_{, l} + p\left[-\frac{2}{K+3}V_{k,k}V_l+V_k V_{k,l}+ V_{l} V_{l, k}\right] \right] \right\}_{,l}+o(\epsilon).
    \end{aligned}
    \end{equation*}
In conclusion, by dropping $o(\epsilon^2)$ terms in Eq. (\ref{cemen}), the Navier-Stokes equations can be derived as follows:
\begin{equation*}
  \begin{aligned}
    \rho_{,t}+(\rho W_k)_{,k}=0,\\
    (\rho V_j)_{,t}+(\rho V_jW_k+p\delta_{jk} -\sigma^\prime_{jk} )_{,k}=-\rho\epsilon_{jlm}\Omega_lV_m,  \\
    \left(\rho E \right)_{,t}+\left(\rho H W_k+pU_k-\kappa T_{,k}-V_l\sigma^\prime_{lk}\right)_{,k} = 0,\\
    \rho E=\frac{1}{2}\rho V_n^2+\frac{K+3}{2}p ,\rho H=\rho E+p ,T=\frac{m}{k}\frac{p}{\rho},
  \end{aligned}
\end{equation*}
where $E$ is the total energy, $H$ is the enthalpy, $T$ is the temperature, $k$ is the Boltzmann constant, $m$ is the mass of a molecule, and $\sigma^\prime_{jk}$ is the stress tensor, which is defined by
\begin{equation*}
  \sigma^\prime_{jk}=\mu(V_{a, l}+V_{l,a}-\frac{2}{3}V_{k,k}\delta_{al})+\beta V_{k,k}\delta_{al} ,
\end{equation*}
where $\mu =\tau p$ is the dynamic viscosity coefficient and
$$\beta=\frac{2}{3}\frac{K}{K+3}\mu$$ is the bulk viscosity coefficient.
The thermal conductivity coefficient $\kappa$ is given by
\begin{equation*}
  \kappa=\frac{K+5}{2}\frac{k}{m}\tau p.
\end{equation*}
In addition, the equations can be written in terms of $\gamma$ instead of $K$ by using $K=(5-3\gamma)/(\gamma-1)$ for 3-Dimensional gas flow. The thermal conductivity becomes
\begin{equation*}
  \kappa=\frac{\gamma}{\gamma-1}\frac{k}{m}\tau p=C_p \mu,
\end{equation*}
where $C_p$ is the specific heat capacity at constant pressure and the Prandtl number is 1.
\section{Moments of the Maxwellian Distribution Function}
\label{moments}
In GKS, the moments of the Maxwellian distribution function with bounded and unbounded integration limits need to be evaluated, and the unbounded integration can refer to \cite{xuGKS2001}. However, when dealing the moving interface, the integration boundary change from 0 to $U$, which means we need to calculate $\int_U^\infty gv^n d\Xi$ and $\int_{-\infty}^U gv^n d\Xi$, denoting as $ <v^{n}>_{v>U} $ and $ <v^{n}>_{v>U} $.
Through the integration by part, the moments are
\begin{equation*}
  \begin{aligned}
    <v^{0}>_{v>U} & = \frac{1}{2} erfc(-\sqrt{\lambda}W),\\
    <v^{1}>_{v>U} & = <v^{0}>_{v>U} + \frac{1}{2\sqrt{\pi\lambda}}e^{-\lambda W^2},\\
    <v^{n+2}>_{v>U} &=V <v^{n+1}>_{v>U} +\frac{n+1}{2\lambda} <v^{n}>_{v>U} +\frac{1}{2\sqrt{\pi\lambda}}U^{n+1} e^{-\lambda W^2},\\
  \end{aligned}
\end{equation*}
and
\begin{equation*}
  \begin{aligned}
    <v^{0}>_{v<U} & = \frac{1}{2} erfc(\sqrt{\lambda}W),\\
    <v^{1}>_{v<U} & = <v^{0}>_{v>U} - \frac{1}{2\sqrt{\pi\lambda}}e^{-\lambda W^2},\\
    <v^{n+2}>_{v>U} &=V <v^{n+1}>_{v>U} +\frac{n+1}{2\lambda} <v^{n}>_{v>U} -\frac{1}{2\sqrt{\pi\lambda}}U^{n+1} e^{-\lambda W^2},\\
  \end{aligned}
\end{equation*}
where $W=V-U$ and $V$ is the macroscopic velocity of fluid element.

\bibliography{mybibfile}

\end{document}